\documentclass[reprint,superscriptaddress,nofootinbib,amsmath,amssymb,aps,prd]{revtex4-2}

\usepackage{graphicx}
\usepackage{dcolumn}
\usepackage{bm}
\usepackage{hyperref}
\usepackage{aas_macros}
\usepackage{tabularx}
\usepackage{multirow} 
\usepackage{physics} 
\usepackage{orcidlink}
\usepackage{soul} 
\usepackage[capitalize]{cleveref}
\usepackage{xcolor}

\newcommand{\logGmu}{\log_{10} G\mu}
\newcommand{\dt}{\vec{\delta t}}
\newcommand{\hyp}{\vec{\eta}}

\begin{document}

\preprint{APS/123-QED}

\title{Practical approaches to analyzing PTA data: Cosmic strings with six pulsars}

\author{Hippolyte Quelquejay Leclere\orcidlink{0000-0002-6766-2004}}
\email{quelquejay@apc.in2p3.fr}
\affiliation{Universit{\'e} Paris Cit{\'e} CNRS, Astroparticule et Cosmologie, 75013 Paris, France}

\author{Pierre Auclair\orcidlink{0000-0002-4814-1406}}
\email{pierre.auclair@uclouvain.be}
\affiliation{Cosmology, Universe and Relativity at Louvain (CURL), Institute of Mathematics and Physics, University of Louvain, 2 Chemin du Cyclotron, 1348 Louvain-la-Neuve, Belgium}

\author{Stanislav Babak\orcidlink{0000-0001-7469-4250}}
\email{stas@apc.in2p3.fr}
\affiliation{Universit{\'e} Paris Cit{\'e} CNRS, Astroparticule et Cosmologie, 75013 Paris, France}

\author{Aurélien Chalumeau\orcidlink{0000-0003-2111-1001}}
\affiliation{Dipartimento di Fisica ``G. Occhialini", Universit{\'a} degli Studi di Milano-Bicocca, Piazza della Scienza 3, I-20126 Milano, Italy}

\author{Dani\`ele A.~Steer\orcidlink{0000-0002-8781-1273}}
\email{steer@apc.in2p3.fr}
\affiliation{Universit{\'e} Paris Cit{\'e} CNRS, Astroparticule et Cosmologie, 75013 Paris, France}

\author{J.~Antoniadis\orcidlink{0000-0003-4453-776}}
\affiliation{Institute of Astrophysics, FORTH, N. Plastira 100, 70013, Heraklion, Greece} 
\affiliation{Max-Planck-Institut f{\"u}r Radioastronomie, Auf dem H{\"u}gel 69, 53121 Bonn, Germany}

\author{A.-S.~Bak~Nielsen\orcidlink{ 0000-0002-1298-9392}}
\affiliation{Max-Planck-Institut f{\"u}r Radioastronomie, Auf dem H{\"u}gel 69, 53121 Bonn, Germany}
\affiliation{Fakult{\"a}t f{\"u}r Physik, Universit{\"a}t Bielefeld, Postfach 100131, 33501 Bielefeld, Germany}

\author{C.~G.~Bassa\orcidlink{0000-0002-1429-9010}}
\affiliation{ASTRON, Netherlands Institute for Radio Astronomy, Oude Hoogeveensedijk 4, 7991 PD, Dwingeloo, The Netherlands}

\author{A.~Berthereau}
\affiliation{Laboratoire de Physique et Chimie de l'Environnement et de l'Espace, Universit\'e d'Orl\'eans / CNRS, 45071 Orl\'eans Cedex 02, France }
\affiliation{Observatoire Radioastronomique de Nan\c{c}ay, Observatoire de Paris, Universit\'e PSL, Université d'Orl\'eans, CNRS, 18330 Nan\c{c}ay, France}

\author{M.~Bonetti\orcidlink{0000-0001-7889-6810}}
\affiliation{Dipartimento di Fisica ``G. Occhialini", Universit{\'a} degli Studi di Milano-Bicocca, Piazza della Scienza 3, I-20126 Milano, Italy}
\affiliation{INFN, Sezione di Milano-Bicocca, Piazza della Scienza 3, I-20126 Milano, Italy}
\affiliation{INAF - Osservatorio Astronomico di Brera, via Brera 20, I-20121 Milano, Italy}

\author{E.~Bortolas}
\affiliation{Dipartimento di Fisica ``G. Occhialini", Universit{\'a} degli Studi di Milano-Bicocca, Piazza della Scienza 3, I-20126 Milano, Italy}
\affiliation{INFN, Sezione di Milano-Bicocca, Piazza della Scienza 3, I-20126 Milano, Italy}
\affiliation{INAF - Osservatorio Astronomico di Brera, via Brera 20, I-20121 Milano, Italy}

\author{P.~R.~Brook\orcidlink{0000-0003-3053-6538}}
\affiliation{Institute for Gravitational Wave Astronomy and School of Physics and Astronomy, University of Birmingham, Edgbaston, Birmingham B15 2TT, UK}

\author{M.~Burgay\orcidlink{0000-0002-8265-4344}}
\affiliation{INAF - Osservatorio Astronomico di Cagliari, via della Scienza 5, 09047 Selargius (CA), Italy}

\author{R.~N.~Caballero\orcidlink{0000-0001-9084-9427}}
\affiliation{Hellenic Open University, School of Science and Technology, 26335 Patras, Greece}

\author{D.~J.~Champion\orcidlink{0000-0003-1361-7723}}
\affiliation{Max-Planck-Institut f{\"u}r Radioastronomie, Auf dem H{\"u}gel 69, 53121 Bonn, Germany}

\author{S.~Chanlaridis\orcidlink{0000-0002-9323-9728}}
\affiliation{Institute of Astrophysics, FORTH, N. Plastira 100, 70013, Heraklion, Greece} 

\author{S.~Chen\orcidlink{0000-0002-3118-5963}}
\affiliation{Kavli Institute for Astronomy and Astrophysics, Peking University, Beijing 100871, P. R. China}

\author{I.~Cognard\orcidlink{0000-0002-1775-9692}}
\affiliation{Laboratoire de Physique et Chimie de l'Environnement et de l'Espace, Universit\'e d'Orl\'eans / CNRS, 45071 Orl\'eans Cedex 02, France }
\affiliation{Observatoire Radioastronomique de Nan\c{c}ay, Observatoire de Paris, Universit\'e PSL, Université d'Orl\'eans, CNRS, 18330 Nan\c{c}ay, France}

\author{G.~Desvignes\orcidlink{0000-0003-3922-4055}}
\affiliation{Max-Planck-Institut f{\"u}r Radioastronomie, Auf dem H{\"u}gel 69, 53121 Bonn, Germany}

\author{M.~Falxa}
\affiliation{Laboratoire de Physique et Chimie de l'Environnement et de l'Espace, Universit\'e d'Orl\'eans / CNRS, 45071 Orl\'eans Cedex 02, France }
\affiliation{Universit{\'e} Paris Cit{\'e} CNRS, Astroparticule et Cosmologie, 75013 Paris, France}

\author{R.~D.~Ferdman}
\affiliation{School of Physics, Faculty of Science, University of East Anglia, Norwich NR4 7TJ, UK}

\author{A.~Franchini\orcidlink{0000-0002-8400-0969}}
\affiliation{Dipartimento di Fisica ``G. Occhialini", Universit{\'a} degli Studi di Milano-Bicocca, Piazza della Scienza 3, I-20126 Milano, Italy}
\affiliation{INFN, Sezione di Milano-Bicocca, Piazza della Scienza 3, I-20126 Milano, Italy}

\author{J.~R.~Gair\orcidlink{0000-0002-1671-3668}}
\affiliation{Max Planck Institute for Gravitational Physics (Albert Einstein Institute), Am Mu{\"u}hlenberg 1, 14476 Potsdam, Germany}

\author{B.~Goncharov\orcidlink{0000-0003-3189-5807}}
\affiliation{Gran Sasso Science Institute (GSSI), I-67100 L'Aquila, Italy }
\affiliation{INFN, Laboratori Nazionali del Gran Sasso, I-67100 Assergi, Italy } 

\author{E.~Graikou}
\affiliation{Max-Planck-Institut f{\"u}r Radioastronomie, Auf dem H{\"u}gel 69, 53121 Bonn, Germany}

\author{J.-M.~Grie{\ss}meier\orcidlink{0000-0003-3362-7996}}
\affiliation{Laboratoire de Physique et Chimie de l'Environnement et de l'Espace, Universit\'e d'Orl\'eans / CNRS, 45071 Orl\'eans Cedex 02, France }
\affiliation{Observatoire Radioastronomique de Nan\c{c}ay, Observatoire de Paris, Universit\'e PSL, Université d'Orl\'eans, CNRS, 18330 Nan\c{c}ay, France}

\author{L.~Guillemot\orcidlink{0000-0002-9049-8716}}
\affiliation{Laboratoire de Physique et Chimie de l'Environnement et de l'Espace, Universit\'e d'Orl\'eans / CNRS, 45071 Orl\'eans Cedex 02, France }
\affiliation{Observatoire Radioastronomique de Nan\c{c}ay, Observatoire de Paris, Universit\'e PSL, Université d'Orl\'eans, CNRS, 18330 Nan\c{c}ay, France}

\author{Y.~J.~Guo}
\affiliation{Max-Planck-Institut f{\"u}r Radioastronomie, Auf dem H{\"u}gel 69, 53121 Bonn, Germany}

\author{H.~Hu\orcidlink{0000-0002-3407-8071}}
\affiliation{Max-Planck-Institut f{\"u}r Radioastronomie, Auf dem H{\"u}gel 69, 53121 Bonn, Germany}

\author{F.~Iraci}
\affiliation{INAF - Osservatorio Astronomico di Cagliari, via della Scienza 5, 09047 Selargius (CA), Italy}
\affiliation{Universit{\'a} di Cagliari, Dipartimento di Fisica, S.P. Monserrato-Sestu Km 0,700 - 09042 Monserrato (CA), Italy}

\author{D.~Izquierdo-Villalba\orcidlink{0000-0002-6143-1491}}
\affiliation{Dipartimento di Fisica ``G. Occhialini", Universit{\'a} degli Studi di Milano-Bicocca, Piazza della Scienza 3, I-20126 Milano, Italy}
\affiliation{INFN, Sezione di Milano-Bicocca, Piazza della Scienza 3, I-20126 Milano, Italy}

\author{J.~Jang\orcidlink{0000-0003-4454-0204}}
\affiliation{Max-Planck-Institut f{\"u}r Radioastronomie, Auf dem H{\"u}gel 69, 53121 Bonn, Germany}

\author{J.~Jawor\orcidlink{0000-0003-3391-0011}}
\affiliation{Max-Planck-Institut f{\"u}r Radioastronomie, Auf dem H{\"u}gel 69, 53121 Bonn, Germany}

\author{G.~H.~Janssen\orcidlink{0000-0003-3068-3677}}
\affiliation{ASTRON, Netherlands Institute for Radio Astronomy, Oude Hoogeveensedijk 4, 7991 PD, Dwingeloo, The Netherlands}
\affiliation{Department of Astrophysics/IMAPP, Radboud University Nijmegen, P.O. Box 9010, 6500 GL Nijmegen, The Netherlands}

\author{A.~Jessner}
\affiliation{Max-Planck-Institut f{\"u}r Radioastronomie, Auf dem H{\"u}gel 69, 53121 Bonn, Germany}

\author{R.~Karuppusamy\orcidlink{0000-0002-5307-2919}}
\affiliation{Max-Planck-Institut f{\"u}r Radioastronomie, Auf dem H{\"u}gel 69, 53121 Bonn, Germany}

\author{E.~F.~Keane\orcidlink{0000-0002-4553-655X}}
\affiliation{School of Physics, Trinity College Dublin, College Green, Dublin 2, D02 PN40, Ireland}

\author{M.~J.~Keith\orcidlink{0000-0001-5567-5492}}
\affiliation{Jodrell Bank Centre for Astrophysics, Department of Physics and Astronomy, University of Manchester, Manchester M13 9PL, UK}

\author{M.~Kramer}
\affiliation{Max-Planck-Institut f{\"u}r Radioastronomie, Auf dem H{\"u}gel 69, 53121 Bonn, Germany}
\affiliation{Jodrell Bank Centre for Astrophysics, Department of Physics and Astronomy, University of Manchester, Manchester M13 9PL, UK}

\author{M.~A.~Krishnakumar\orcidlink{0000-0003-4528-2745}}
\affiliation{Max-Planck-Institut f{\"u}r Radioastronomie, Auf dem H{\"u}gel 69, 53121 Bonn, Germany}
\affiliation{Fakult{\"a}t f{\"u}r Physik, Universit{\"a}t Bielefeld, Postfach 100131, 33501 Bielefeld, Germany}

\author{K.~Lackeos\orcidlink{0000-0002-6554-3722}}
\affiliation{Max-Planck-Institut f{\"u}r Radioastronomie, Auf dem H{\"u}gel 69, 53121 Bonn, Germany}

\author{K.~J.~Lee}
\affiliation{Institute of Astrophysics, FORTH, N. Plastira 100, 70013, Heraklion, Greece} 
\affiliation{Max-Planck-Institut f{\"u}r Radioastronomie, Auf dem H{\"u}gel 69, 53121 Bonn, Germany}
\affiliation{Observatoire Radioastronomique de Nan\c{c}ay, Observatoire de Paris, Universit\'e PSL, Université d'Orl\'eans, CNRS, 18330 Nan\c{c}ay, France}

\author{K.~Liu}
\affiliation{Max-Planck-Institut f{\"u}r Radioastronomie, Auf dem H{\"u}gel 69, 53121 Bonn, Germany}

\author{Y.~Liu\orcidlink{0000-0001-9986-9360}}
\affiliation{Fakult{\"a}t f{\"u}r Physik, Universit{\"a}t Bielefeld, Postfach 100131, 33501 Bielefeld, Germany}
\affiliation{National Astronomical Observatories, Chinese Academy of Sciences, Beijing 100101, P. R. China}

\author{A.~G.~Lyne}
\affiliation{Jodrell Bank Centre for Astrophysics, Department of Physics and Astronomy, University of Manchester, Manchester M13 9PL, UK}

\author{J.~W.~McKee\orcidlink{0000-0002-2885-8485}}
\affiliation{E.A. Milne Centre for Astrophysics, University of Hull, Cottingham Road, Kingston-upon-Hull, HU6 7RX, UK}
\affiliation{Centre of Excellence for Data Science, Artificial Intelligence and Modelling (DAIM), University of Hull, Cottingham Road, Kingston-upon-Hull, HU6 7RX, UK}

\author{R.~A.~Main}
\affiliation{Max-Planck-Institut f{\"u}r Radioastronomie, Auf dem H{\"u}gel 69, 53121 Bonn, Germany}

\author{M.~B.~Mickaliger\orcidlink{0000-0001-6798-5682}}
\affiliation{Jodrell Bank Centre for Astrophysics, Department of Physics and Astronomy, University of Manchester, Manchester M13 9PL, UK}

\author{I.~C.~Ni\c{t}u\orcidlink{0000-0003-3611-3464}}
\affiliation{Jodrell Bank Centre for Astrophysics, Department of Physics and Astronomy, University of Manchester, Manchester M13 9PL, UK}

\author{A.~Parthasarathy\orcidlink{0000-0002-4140-5616}}
\affiliation{Max-Planck-Institut f{\"u}r Radioastronomie, Auf dem H{\"u}gel 69, 53121 Bonn, Germany}

\author{B.~B.~P.~Perera\orcidlink{0000-0002-8509-5947}}
\affiliation{Arecibo Observatory, HC3 Box 53995, Arecibo, PR 00612, USA}

\author{D.~Perrodin\orcidlink{0000-0002-1806-2483}}
\affiliation{INAF - Osservatorio Astronomico di Cagliari, via della Scienza 5, 09047 Selargius (CA), Italy}

\author{A.~Petiteau\orcidlink{0000-0002-7371-9695}}
\affiliation{IRFU, CEA, Université Paris-Saclay, F-91191 Gif-sur-Yvette, France}
\affiliation{Universit{\'e} Paris Cit{\'e} CNRS, Astroparticule et Cosmologie, 75013 Paris, France}

\author{N.~K.~Porayko}
\affiliation{Max-Planck-Institut f{\"u}r Radioastronomie, Auf dem H{\"u}gel 69, 53121 Bonn, Germany}
\affiliation{Dipartimento di Fisica ``G. Occhialini", Universit{\'a} degli Studi di Milano-Bicocca, Piazza della Scienza 3, I-20126 Milano, Italy}

\author{A.~Possenti}
\affiliation{INAF - Osservatorio Astronomico di Cagliari, via della Scienza 5, 09047 Selargius (CA), Italy}

\author{A.~Samajdar\orcidlink{0000-0002-0857-6018}}
\affiliation{Institut f\"{u}r Physik und Astronomie, Universit\"{a}t Potsdam, Haus 28, Karl-Liebknecht-Str. 24/25, 14476, Potsdam, Germany}

\author{S.~A.~Sanidas}
\affiliation{Jodrell Bank Centre for Astrophysics, Department of Physics and Astronomy, University of Manchester, Manchester M13 9PL, UK}

\author{A.~Sesana}
\affiliation{Dipartimento di Fisica ``G. Occhialini", Universit{\'a} degli Studi di Milano-Bicocca, Piazza della Scienza 3, I-20126 Milano, Italy}
\affiliation{INFN, Sezione di Milano-Bicocca, Piazza della Scienza 3, I-20126 Milano, Italy}
\affiliation{INAF - Osservatorio Astronomico di Brera, via Brera 20, I-20121 Milano, Italy}

\author{G.~Shaifullah\orcidlink{0000-0002-8452-4834}}
\affiliation{Dipartimento di Fisica ``G. Occhialini", Universit{\'a} degli Studi di Milano-Bicocca, Piazza della Scienza 3, I-20126 Milano, Italy}
\affiliation{INFN, Sezione di Milano-Bicocca, Piazza della Scienza 3, I-20126 Milano, Italy}
\affiliation{INAF - Osservatorio Astronomico di Cagliari, via della Scienza 5, 09047 Selargius (CA), Italy}

\author{L.~Speri\orcidlink{0000-0002-5442-7267}}
\affiliation{Max Planck Institute for Gravitational Physics (Albert Einstein Institute), Am Mu{\"u}hlenberg 1, 14476 Potsdam, Germany}

\author{R.~Spiewak}
\affiliation{Jodrell Bank Centre for Astrophysics, Department of Physics and Astronomy, University of Manchester, Manchester M13 9PL, UK}

\author{B.~W.~Stappers}
\affiliation{Jodrell Bank Centre for Astrophysics, Department of Physics and Astronomy, University of Manchester, Manchester M13 9PL, UK}

\author{S.~C.~Susarla\orcidlink{0000-0003-4332-8201}}
\affiliation{Ollscoil na Gaillimhe --- University of Galway, University Road, Galway, H91 TK33, Ireland}

\author{G.~Theureau\orcidlink{0000-0002-3649-276X}}
\affiliation{Laboratoire de Physique et Chimie de l'Environnement et de l'Espace, Universit\'e d'Orl\'eans / CNRS, 45071 Orl\'eans Cedex 02, France }
\affiliation{Observatoire Radioastronomique de Nan\c{c}ay, Observatoire de Paris, Universit\'e PSL, Université d'Orl\'eans, CNRS, 18330 Nan\c{c}ay, France}
\affiliation{Laboratoire Univers et Th{\'e}ories LUTh, Observatoire de Paris, Universit{\'e} PSL, CNRS, Universit{\'e} de Paris, 92190 Meudon, France}

\author{C.~Tiburzi}
\affiliation{INAF - Osservatorio Astronomico di Cagliari, via della Scienza 5, 09047 Selargius (CA), Italy}

\author{E.~van~der~Wateren\orcidlink{0000-0003-0382-8463}}
\affiliation{Department of Astrophysics/IMAPP, Radboud University Nijmegen, P.O. Box 9010, 6500 GL Nijmegen, The Netherlands}
\affiliation{ASTRON, Netherlands Institute for Radio Astronomy, Oude Hoogeveensedijk 4, 7991 PD, Dwingeloo, The Netherlands}

\author{A.~Vecchio\orcidlink{0000-0002-6254-1617}}
\affiliation{Institute for Gravitational Wave Astronomy and School of Physics and Astronomy, University of Birmingham, Edgbaston, Birmingham B15 2TT, UK}

\author{V.~Venkatraman~Krishnan\orcidlink{0000-0001-9518-9819}}
\affiliation{Max-Planck-Institut f{\"u}r Radioastronomie, Auf dem H{\"u}gel 69, 53121 Bonn, Germany}


\author{J.~Wang\orcidlink{0000-0003-1933-6498}}
\affiliation{Fakult{\"a}t f{\"u}r Physik, Universit{\"a}t Bielefeld, Postfach 100131, 33501 Bielefeld, Germany}
\affiliation{Ruhr University Bochum, Faculty of Physics and Astronomy, Astronomical Institute (AIRUB), 44780 Bochum, Germany}
\affiliation{Advanced Institute of Natural Sciences, Beijing Normal University, Zhuhai 519087, China }

\author{L.~Wang}
\affiliation{Jodrell Bank Centre for Astrophysics, Department of Physics and Astronomy, University of Manchester, Manchester M13 9PL, UK}

\author{Z.~Wu\orcidlink{0000-0002-1381-7859}}
\affiliation{National Astronomical Observatories, Chinese Academy of Sciences, Beijing 100101, P. R. China}

\collaboration{The European Pulsar Timing Array}

\date{\today}

\begin{abstract}
We search for a stochastic gravitational wave background (SGWB) generated by a network of cosmic strings using six millisecond pulsars from Data Release 2 (DR2) of the European Pulsar Timing Array (EPTA). 
We perform a Bayesian analysis considering two models for the network of cosmic string loops, and compare it to a simple power-law model which is expected from the population of supermassive black hole binaries. Our main strong assumption is that the previously reported common red noise process is a SGWB.
We find that the one-parameter cosmic string model is slightly favored over a power-law model thanks to its simplicity. If we assume a two-component stochastic signal in the data (supermassive black hole binary population and the signal from cosmic strings), we get a $95\%$ upper limit on the string tension of $\log_{10}(G\mu) < -9.9$ ($-10.5$) for the two cosmic string models we consider.  In  extended two-parameter string models, we were unable to constrain the number of kinks.
We test two approximate and fast Bayesian data analysis methods against the most rigorous analysis and find consistent results. 
These two fast and efficient methods are applicable to all SGWBs, independent of their source, and will be crucial for analysis of extended data sets.
\end{abstract}

\maketitle

\section{Introduction}

All regional Pulsar Timing Array (PTA) collaborations have recently, and independently, reported the presence of a common red noise process in the observed data \cite{EPTA_CRS, NANOGravCRS, PPTA_CRS}. Furthermore, the combined dataset shows even stronger evidence for its presence \cite{IPTA_CRS}. The exact nature of this signal is as yet very uncertain. While its spectral properties are consistent with an expected stochastic gravitational wave background (SGWB), the data is not sensitive enough to make any informative statement for or against its gravitational wave (GW) nature \footnote{After submission of this paper on the ArXiV and during the peer review process, the results from the regional PTA collaborations became public \cite{EPTA:2023fyk, Agazie_2023, PPTA_2023ApJ, Xu:2023wog, NANOGrav:2023hfp}. These papers have demonstrated evidence of GW-induced correlations in their respective datasets
with a considerable statistical significance.}.

In this paper, we make a \emph{strong} assumption that it is, in fact, a SGWB and work towards its possible interpretation. The most favorable model for an anticipated SGWB is a superposition of monochromatic GW signals from a population of supermassive black hole binaries (SMBHBs) in the local Universe \cite{EPTA_CRS, Sesana_2013}.  This signal, assuming SMBHBs in circular orbits, is a power-law with theoretical spectral index $\gamma = 13/3$ in the power spectral density of the residuals \cite{phinney2001practical}. However, realistic astrophysical simulations suggest that the spectral index could vary based on the realization of the observed Universe \cite{Sesana_2013_gwe}. For this reason, the spectral index is usually inferred from the data, giving a two-parameter model: $A, \gamma$. The amplitude $A$ is referenced at the frequency 1/year.

We propose an alternative to SMBHBs, namely a SGWB from an early universe source, and in particular a network of cosmic string loops. (See \cite{Ellis:2020ena,Blasi:2020mfx,Bian:2022tju,Chen:2022azo, Hindmarsh_2023} for previous work on cosmic strings in the nHz band.) Cosmic strings are topological defects that could have emerged from symmetry-breaking phase transitions in the early Universe~\cite{Nielsen:1973cs,Kibble:1976sj,Kibble:1984hp,Jeannerot:2003qv}.
These quasi one-dimensional objects are characterized by their dimensionless tension $G\mu$, where $G$ is Newton's constant.
Numerical simulations show that on large scales, cosmic string networks reach an attractor `scaling' regime in which all the characteristic length scales in the network grow as $t$ \cite{Ringeval:2005kr,Martins:2005es,Olum:2006ix,Hindmarsh:2008dw}.
Cosmic string loops are formed at all times by the scaling infinite string network, oscillate with period $\ell/2$, and decay into GWs\footnote{Field theory simulations show that cosmic string loops also decay into particles~\cite{Matsunami:2019fss,Hindmarsh:2021mnl}. Although the balance between GW and particle emission is still under debate, preliminary quantitative studies tend to show that the SGWB from cosmic string is unaffected at the frequency of PTAs~\cite{Auclair:2019jip,Auclair:2021jud}}.
The superposition of the GW emitted generates a SGWB that depends on the cosmic string loop distribution.
While the loop distribution is well-understood on large scales close to the Hubble radius, there is an active debate regarding the distribution of smaller loops. 
To account for this theoretical uncertainty, in the following we consider two loop distribution models that have been used by the LIGO-Virgo-Kagra \cite{LIGOScientific:2017ikf,LIGOScientific:2021nrg}
 and LISA \cite{Auclair_2020,LISACosmologyWorkingGroup:2022jok} collaborations.

In this paper we use the six pulsars (see \autoref{tab:pulsar models}) from the early Data Release 2 (DR2) of European PTA (EPTA) collaboration \cite{EPTA_CRS} \footnote{This dataset is a subset of the full DR2 dataset recently used in \cite{EPTA:2023fyk} which includes a total of 25 pulsars.}
Six pulsars, at the current sensitivity and observational time span, are not sufficient to detect the stochastic GW signal, however the presence of the common red process is evident. More pulsars are required to populate the expected Hellings-Downs (HD) \cite{Hellings1983UpperLO} correlation curve and to get a high statistical significance for a SGWB. Nevertheless, in our study, we still utilize Hellings-Downs correlations to analyze the common red noise among pulsars. This approach allows us to account for the correlations among pulsars when characterizing the common red noise (CRN), resulting in a distinct interaction with the individual red noises.
Given that we consider only 6 pulsars, the diagonal noise component (autocorrelation part of the noise matrix) will dominate and we expect only a small difference from the inclusion of  Hellings-Downs correlation. Here we describe in detail the methodology of inferring parameters of the string network. This same methodology will later be applied on the extended dataset~\cite{EPTA:2023fyk}.

Assuming that the observed common red noise is a SGWB, we use Bayesian methods to infer the characteristics of the string network using a one-parameter model ($G\mu$) and a two-parameter model ($G\mu,N_k$) where $N_k$ is the average number of kinks on a loop per oscillation period, see \autoref{sec:SGWB_part}. 
In addition to the SGWB signal, we also model the three components of the noise: white noise, pulsar red noise and chromatic noises (dispersion measurement variations and scattering variations) (see \cite{Chalumeau_2021} for details).

We carry the Bayesian analysis through three different approaches with varying accuracy of description and computational cost.
The first approach, dubbed in this paper as ``Full method'' \cite{van_Haasteren_2012, van_Haasteren_2014}, is the standard approach in which the parameters of the SGWB and of the noise are explored simultaneously. This method is the most accurate but also the most computationally expensive due to the high-dimensionality of the analysis and because we consider Hellings-Downs correlations for the SGWB.
The second approach, dubbed as ``Resampling method'' \cite{Hourihane:2022ner}, neglects at first the spatial correlation between pulsars considering a common uncorrelated signal. The resulting posterior distributions are then resampled by taking into account GW-induced correlation. In this approach, the likelihood is factorized into a product of likelihoods for each pulsar \cite{Taylor_2022}, and the computational cost is significantly reduced with respect to the Full method as the number of pulsars increases.
Finally, the third method, dubbed as ``Free spectrum'' \cite{Taylor_NFS, Moore_2021}, amounts to obtaining the correlated power of the common noise for each frequency bin independently, marginalizing over the single pulsar noise parameters, \emph{before} inferring the SGWB parameters.
This method drastically reduces the dimensionality of the analysis and allows for a very fast parameter estimation of the SGWB models.

The paper is organized as follows. We start with a description of the SGWB produced by cosmic string loops in \autoref{sec:SGWB_part}.
In \autoref{sec:DataAnalysis} we explain the three data analysis approaches used.
In \autoref{sec:results}, we present the constraints on single parameter models $(G\mu)$ and two parameter models $(G\mu, N_k)$.
We perform a Bayesian model comparison between the different cosmic string models and a simple power-law SGWB. 
Finally, we conclude in \autoref{sec:concl}.

\section{Stochastic Gravitational Wave Background from cosmic strings}
\label{sec:SGWB_part}

The SGWB generated by a cosmic string network has been studied in depth, see e.g.~\cite{Blanco_Pillado_2014,Auclair_2020}. It includes a contribution from the uncorrelated superposition GW bursts emitted from cusps, kinks and kink-kink collisions on oscillating loops \cite{Damour_2001}.
Cusps, which are points at which the loop reaches ultra relativistic velocities, emit a short beam of GWs. 

Kinks, which are discontinuities in the tangent vector of the loop, propagate along the string at the speed of light and emit a one-dimensional ``fan-like'' GW burst with the same beaming angle as for cusps \cite{Damour_2001}. 
Finally, when left- and right-moving kinks collide (kink-kink collisions), an isotropic burst of GW emission is emitted \cite{Ringeval_2017}.

A loop of length $\ell$ oscillates periodically with corresponding fundamental frequency $f_0 = 2 / \ell$.
For the $n$-th oscillatory mode of the loop, the power (in units of $G\mu^2$) emitted in GWs by each of these burst events (labelled by $b$) is given by
\footnote{Formally this expression is only valid for large $n$, but as elsewhere we assume that extrapolating to $n \geq 1$ is a good approximation to describe low-harmonic modes.}
\cite{Damour_2001}
\begin{equation}
    \label{Power emitted by the n-th mode function of i}
    P_n^{(b)} = \frac{\Gamma^{(b)}}{\zeta(q_b)}n^{-q_b}.
\end{equation}

Here $\zeta(q_b)$ is the Riemann zeta function, $q_b=4/3$, $5/3$ and $2$ for cusps, kinks and kink-kink collisions respectively and \cite{Damour_2001,Ringeval_2017}
\begin{equation}
    \label{different gamma values for each events}
    \Gamma^c = \frac{3(\pi g_1^c)^2}{2^{1/3}g_2^{2/3}}, \: \Gamma^k = \frac{3(\pi g_1^k)^2}{2^{2/3}g_2^{1/3}}, \: \Gamma^{kk} = 2(\pi g_1^{kk})^2 \;,
\end{equation}
where $g_2 = \sqrt{3}/4, g_1^c\approx 0.85, g_1^k\approx 0.29, g_1^{kk}\approx 0.1$.

Given the exponents $q_b$ in \cref{Power emitted by the n-th mode function of i}, one might expect that, compared to the cusp contribution, the kink and kink-kink collision contributions could be neglected. However, as already noted in \cite{Ringeval_2017},
if many kinks are present they can dominate the power emitted in GWs.
Indeed, since two kinks (right and left moving) are formed when two pieces of strings intercommute, the number of kink-kink collisions on a loop with $N_k$ kinks per oscillation period should scale as $N_k^2/4$, which can then dominate for high $N_k$.
Furthermore, kinks can proliferate on loops with junctions~\cite{Binetruy:2010bq,Binetruy:2010cc}.
To summarize, for a loop containing $N_c$ cusps and $N_k$ kinks per oscillation period, the total power emitted, in units of $G\mu^2$, is
\begin{equation}
    \label{eq: total Gamma}
    \Gamma = N_c\Gamma^{c} + N_k\Gamma^{k} + \frac{N_k^2}{4} \Gamma^{kk}.
\end{equation}

The fractional energy density of the SGWB per logarithmic interval of frequency is
\begin{equation}
    \label{Loop energy density fraction}
    \Omega_{gw}(t_0, f) \equiv \frac{8\pi G}{3H_0^2} f \dv{\rho_{gw}}{f} \: ,
\end{equation}
where $H_0$ is the Hubble constant, and $\dv*{\rho_{gw}}{f}$ is the energy density in gravitational waves per unit frequency $f$, observed today (at $t = t_0$). Following \cite{Vilenkin1995CosmicSA,LIGOScientific:2021nrg}, for a network of cosmic string loops with distribution $\textbf{n}(\ell,t)$, this is given by
\begin{multline}
    \label{eq: full SGWB energy density}
    \dv{\rho_{gw}}{\ln f} =  \frac{2 G\mu^2}{f} \sum_b \frac{N_b \Gamma^{(b)}}{\zeta(q_b)} \\ 
    \times \sum_{n=1}^{+\infty} \int \frac{n^{1-q_b} \dd{z}}{(1+z)^6 H(z)}
\textbf{n}\left[\frac{2n}{(1+z)f}, t(z)\right],
\end{multline}
where we have summed over the cusp, kink and kink-kink contributions, and $H(z)$ is the Hubble parameter. In the following we consider standard $\Lambda$CDM cosmology with the Planck-2018 fiducial parameters \cite{Planck:2018vyg}.

\begin{figure}
\includegraphics[width=0.48\textwidth]{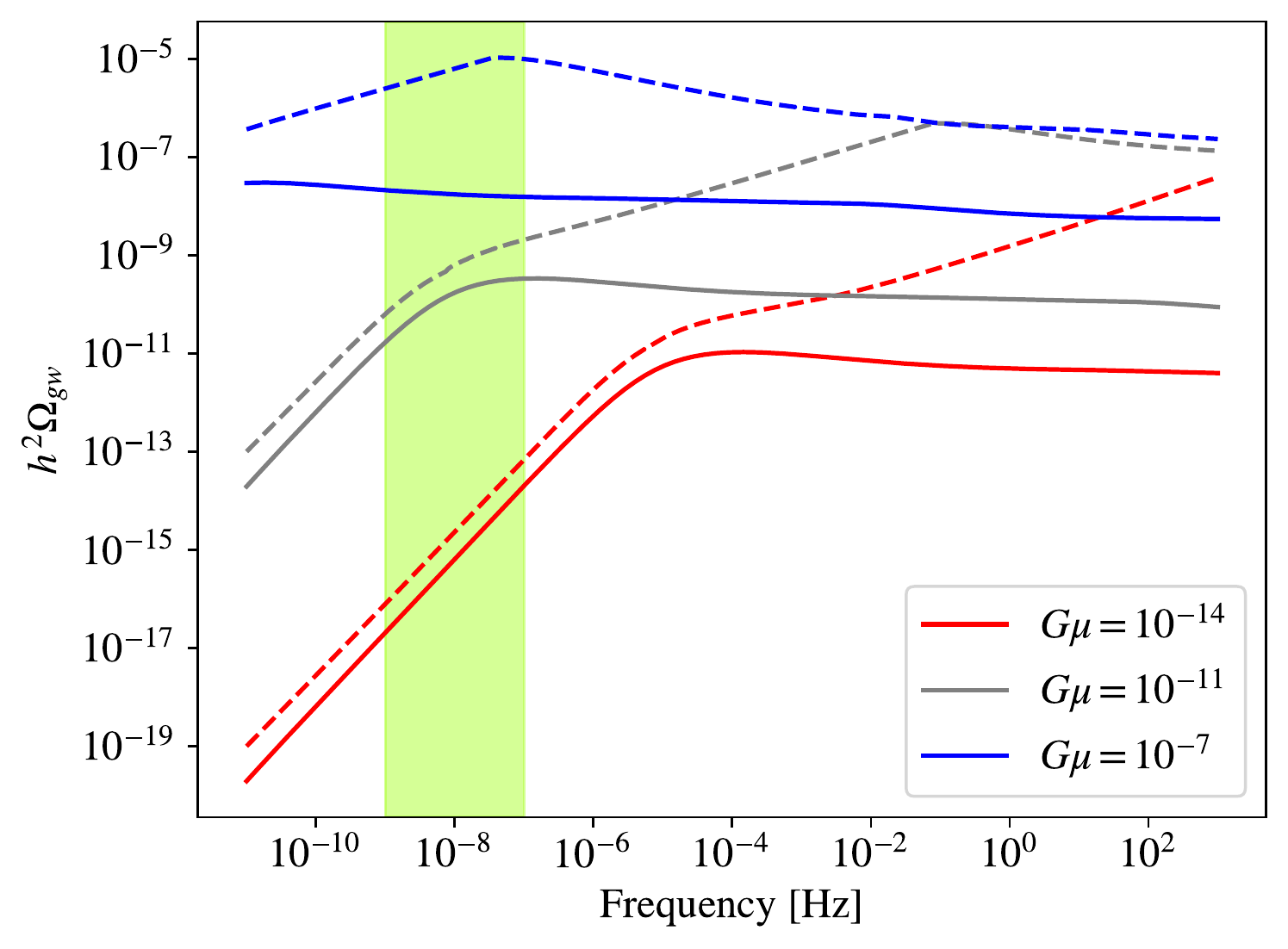}
\caption{SGWB from a network of cosmic string loops, expressed in terms of characteristic energy density. The spectrum is computed using
the BOS (resp.~LRS) loop number density model for the solid (resp.~dashed) lines. Here we have taken $N_c=2, N_k=0$ leading to $\Gamma =57$. For each model, computations using three different tension values $G\mu$ are represented. The sensitivity frequency range of EPTA corresponds to the yellow band.}
\label{fig:SGWB_wide}
\end{figure}

We consider the most updated loop distribution models $\mathbf{n}(\ell,t)$ available, calibrated with numerical simulations, that have also been used in LVK collaboration papers \cite{LIGOScientific:2017ikf, LIGOScientific:2021nrg} and LISA~\cite{Auclair:2019wcv,LISACosmologyWorkingGroup:2022jok}. They are denoted in the following as BOS \cite{Blanco_Pillado_2014, Blanco_Pillado_2017} and LRS \cite{Lorenz_Ringeval_2010}. Each has an intersection probability $p=1$, and the largest loops produced are of size $0.1 t$.    The main difference between these models is that, relative to the BOS model, the LRS model has an additional population of very small loops which aim to account for physics at the gravitational backreaction scale. These small loops emit high frequency GWs, and lead to modifications of the SGWB at high frequencies $f \gg H_0 (\Gamma G\mu)^{-1}$ \cite{Auclair:2020oww}.
This can be seen in \autoref{fig:SGWB_wide} which plots the numerical evaluation of \cref{eq: full SGWB energy density} for $N_c=2$, $N_k=0$ for the BOS model (continuous lines) and LRS model (dashed lines) for different values of $G\mu$.

Notice that as a consequence, the two models considered give a significantly different spectrum in PTA frequency range only for $\Gamma G\mu \gtrsim 2 \times 10^{-11}$ (as seen for $G\mu = 10^{-14}$ and $\Gamma = 57$ in \autoref{fig:SGWB_wide}).

In this work, we consider two cases (i) smooth cosmic string loops with two cusps only (as in \autoref{fig:SGWB_wide}), (ii) fixed number of cusps, $N_c=1$, but a varying number of kinks $N_k$ with $N_k^2/4$ kink-kink collisions.
Our aim is to quantify how well the SGWB from cosmic string loops can explain the common red process that is already seen with strong evidence in three PTAs, namely the EPTA, NANOGRAV and PPTA consortiums \cite{EPTA_CRS, NANOGravCRS, PPTA_CRS}.

\section{\label{sec:DataAnalysis}Data analysis method}

In this section, we use a Bayesian analysis to perform parameter estimation and model selection.  We will not repeat the basis of Bayesian inference in the context of PTA, and instead refer the reader to \cite{EPTA_CRS, Chalumeau_2021} for a more detailed description. In the following subsection, we briefly describe the noise and SGWB models which we test.

We have applied three Bayesian approaches to the six pulsars of the early second EPTA data release (DR2)~\cite{EPTA_CRS}:
\begin{enumerate}
\item  We directly infer the parameters of each model using the spectral shape of the SGWB as a prior, considering HD correlation between pulsars. This is the most computationally expensive path, which is referred to as ``full''.
\item Similar to 1, but initially we neglect all correlations between pulsars, assuming an uncorrelated common process. Then, we resample the posterior points using ``correct'' likelihood \emph{with} HD correlation following the procedure outlined in \cite{Hourihane:2022ner}. This path is significantly faster, we refer to it as ``resampling'' (RS).
\item Finally, we also use the method suggested in \cite{Moore_2021} and further developed in \cite{Taylor_NFS}. There, we first estimate the correlated power at each Fourier bin with a relatively large log-uniform prior: we refer to this as the ``free spectrum''. Then, we perform Bayesian analysis using the free spectrum as observational data. This is the computationally cheapest path, which we refer to as the ``free spectrum'' (FS) method. 
\end{enumerate}

\subsection{PTA data analysis principles}

In PTA, we work with time residuals, obtained as the difference between observed and predicted time of arrivals of the pulsars' radio pulses
\begin{equation}
    \label{eq: def of residuals}
    \delta t_i = t^{\rm{obs}}_i - t^{\rm{TM}}_i (\vec{\beta_{b}}),
\end{equation}
where $i$ is the index of each observation and $t^{\rm{TM}}$ is the timing model attempting to explain the deviations from the observed time-of-arrival using a set of parameters $\vec{\beta_{b}}$. Following the usual path \cite{van_Haasteren_2012, van_Haasteren_2014} we marginalize analytically over timing model parameters.

We assume that the noise in each pulsar consists of the radiometer white noise, intrinsic pulsar spin noise, dispersion measurement and finally scattering variations. These last two components are referred to as chromatic noises, since they are due to the propagation of the radio pulses through the interstellar medium and depend on the observational radio frequency.  We follow the standard procedure \cite{Chalumeau_2021}, initially we infer the noise model in each pulsar individually, and then we use these results as a starting point  in the search for the SGWB common to all pulsars.

Besides these individual noise components, we also introduce SGWB into the data model. It is characterized by the spectral shape (subject to its source) and exhibits a particular correlation across pulsar pairs which depends on the angular separation of pulsars in the sky and is described by the Hellings-Downs (HD) curve \cite{Hellings1983UpperLO}. 

We assume that the data does not contain any deterministic signals, with the exception of the two deterministic chromatic signals, referred to as exponential dips, of PSR J1713+0747 (see \autoref{tab:pulsar models}).

We can summarize the data model in the form of the total covariance matrix
\begin{equation}
    \label{eq:full_hierar_cov_matrix}
    C_{(ai)(bj)} = \mathcal{N}_{a,(ij)}\delta_{ab} + (C^{\rm{RN}}_{a,(ij)} + C^{\rm{ChRN}}_{a,(ij)})\delta_{ab} + \Gamma_{ab}C^{\rm{CRN}}_{(ij)},
\end{equation}
where $\delta_{(..)}$ is the Kronecker delta function, $a, b$ are indexing the pulsars, $i, j$ the timing residuals, $\mathcal{N}_a$ is the white noise covariance matrix of pulsar $a$, $C^{\rm{RN}}_a$ its intrinsic red-noise covariance matrix and $C^{\rm{ChRN}}_a$ its chromatic red-noise covariance matrix. $C^{\rm{CRN}}$ is the common red noise covariance matrix, and $\Gamma_{ab}$ is the overlap reduction function giving the correlations between pulsars $a$ and $b$. The function $\Gamma_{ab}$ reduces to $\delta_{ab}$ for an uncorrelated common red noise and is described by the HD correlations for a SGWB, 
\begin{equation}
    \label{eq: Hellings down curve}
    \Gamma_{ab} = \frac{3}{2}x\ln x - \frac{1}{4}x + \frac{1}{2} (1 + \delta_{ab}),
\end{equation}
where $x = \frac{1 - \cos \xi_{ab}}{2}$, $\xi_{ab}$ being the angular separation between pulsars $a$ and $b$.

In PTA data analysis, red stochastic processes are often described via Gaussian Processes with truncated Fourier basis (see \cite{van_Haasteren_2014} for details)
\begin{equation}
    \label{eq:Red noise realization}
    \delta t_{\rm{red}}(t_i) = \sum_{n=1}^{N_f} \left[ a_n \cos \left( \frac{2\pi n t_i}{T} \right) + b_n \sin \left( \frac{2\pi n t_i}{T} \right)  \right],
\end{equation}
 where $T$ is the observational time span and $a_n, b_n$ are random variables (weights) defined by the spectral shape of the underlying process.  In this work, we use $N_f = 30$ for the common red process while for the individual pulsar noises, we employ a custom number of Fourier bins determined through single pulsar analysis, following the method detailed in \cite{Chalumeau_2021} (see \autoref{tab:pulsar models} for an overview of the different $N_f$ values used for each pulsar noise). For the chromatic noises, we use a similar approach but introduce a scaling with the observational radio frequency $\nu$, $\delta t_{\rm{red}} \propto \nu^{-\chi}$, where $\chi$ is a chromatic index equals to $2$ for dispersion measure variations and $4$ for scattering variations.
 
For all intrinsic red noise components attached to each pulsar, we assume a power-law shape for the one-sided power spectral density (PSD) of the residuals
\begin{equation}
    \label{eq: powerlaw PSD}
     S(f; A, \gamma) = \frac{A^2}{12\pi^2} \left( \frac{f}{\text{yr}^{-1}} \right)^{-\gamma} \text{yr}^3 \: ,
\end{equation}
characterized by the spectral index $\gamma$ and the amplitude $A$ defined at the reference frequency of $1/$year.

\begin{table*}
    \renewcommand{\arraystretch}{1.5}
    \begin{ruledtabular}
        \begin{tabular}{ccccccc}
        
            Pulsar & Time Span (years) & Number of ToAs & Red Noise & DM Variations & Scattering Variations & Deterministic Signals \\        
            J0613$-$0200 & 22.89 & 2909 & 10 & 144 & $-$ & $-$ \\
            J1012$+$5307 & 23.68 & 5325 & 149 & 45 & $-$ & $-$ \\
            J1600$-$3053 & 14.32 & 2982 & $-$ & 26 & 137 & $-$ \\
            J1713$+$0747 & 24.46 & 5003 & 11 & 148 & $-$ & Two exponential dips\footnote{Interested readers are directed to section 5.2.2 of \cite{Chalumeau_2021} for further details regarding these events.} \\ 
            J1744$-$1134 & 24.01 & 1946 & 9 & 151 & $-$ & $-$ \\ 
            J1909$-$3744 & 15.74 & 2503 & 20 & 151 & $-$ & $-$ \\
        \end{tabular}
    \end{ruledtabular}
    \caption{Noise model used for each of six pulsars. For each pulsar, we also consider white noise whose parameters are set to the maximum likelihood values obtained through single pulsar analysis. We indicate the number of Fourier bins ($N_f$ in \autoref{eq:Red noise realization}) employed for each noise type and pulsar. In cases where a particular noise type is not included in the pulsar's noise model, we denote it with a $-$ sign.}
    \label{tab:pulsar models}
\end{table*}

Here we assume that the CRN is a SGWB with the one-sided power spectral density given in terms of the fractional energy density by
\begin{equation}
    S(f ; \hyp) = \frac{H_0^2}{8\pi^4}\frac{1}{f^5} \: \Omega_{\rm{SGWB}}(f ; \hyp),
    \label{eq:StoOmegaGW}
\end{equation}
where $\hyp$ are the hyper-parameters of the SGWB model considered. The SGWB from a population of SMBHB is well approximated by a power-law model, and takes the same form as in \eqref{eq: powerlaw PSD}, see \cite{Sesana_2013}.  Moreover, for the SMBHBs in circular orbits with GW-driven evolution, we expect $\gamma \approx 13/3$. 
The spectral shape for the network of cosmic strings is obtained by integration of \cref{eq: full SGWB energy density} which adds computational time.

In the following analysis, we fix the parameters (and, therefore the level) of the white noise based on each pulsar investigation \cite{Chalumeau_2021}. We vary the parameters of all red noise components together with the parameters of SGWB. In total, there are 30 + $N_{\rm{CRN}}$ parameters, where $N_{\rm{CRN}}$ is the number of hyper-parameters of the CRN (varies from model to model).

\subsection{``Full'' Method}
\label{sec:classicMeth}

This is the standard path one would follow to infer the parameters of each model (pulsars and SGWB). The likelihood for the concatenated array of observations (residuals) $\vec{\delta t}$ is given by the usual Gaussian form
\begin{equation}
    \label{eq:full_hierar_lklhd}
    p(\dt|\hyp) = \frac{1}{\sqrt{|2\pi C|}} \exp\left( -\frac{1}{2} \dt^T C^{-1} \dt \right),
\end{equation}
where $C$ is the full covariance matrix of \cref{eq:full_hierar_cov_matrix}. 
It is parametrized by $A_a, \gamma_a$ for each red noise component for each pulsar $a$ and by parameters describing the SGWB spectrum. We use the python package \texttt{ENTERPRISE} \cite{ENTERPRISE} for its computation (with an extension to include cosmic string models). We sample parameters using \texttt{PTMCMC} sampler \cite{ptmcmc}. Note that we do not employ the ``parallel tempering'' features due to technical reasons.

For the SGWB we have used either the power-law model, the BOS or the LRS models with fixed and variable number of kinks. This is computationally the most expensive (but most rigorous) approach. There are two bottlenecks in the analysis: (i) integration of \cref{eq: full SGWB energy density}, and (ii) accounting for the correlation between pulsars (inverting the $\Gamma_{ab}$ part in the covariance matrix). However, thanks to the low dimensionality of the parameter space of the cosmic string spectrum ($f$, $G\mu$, $N_k$), we were able to address point (i) by using numerical interpolation method over a precomputed grid to speed-up the computation of its power spectrum.

We can also perform Bayesian model selection. We use the Bayes Factor (BF) as a way to measure which model is preferred by the data. The Bayes Factor is equivalent to the posterior odds ratio $\mathcal{O}$ if we assume (and we do) equal model priors
\begin{equation}
    \label{eq: BF_odd_ratio}
    \mathcal{B}^{M_1}_{M_2} = \frac{p(d|M_1)}{p(d|M_2)} = \frac{p(M_1 | d)}{p(M_2 | d)}\frac{p(M_2)}{p(M_1)} = \mathcal{O}^{M_1}_{M_2}\frac{p(M_2)}{p(M_1)}.
\end{equation}

It is computationally prohibitive to compute the evidence for each model directly using the Nested sampling method. Instead, we used the product-space sampling approach described in \cite{Hee_product} and already used in several PTA analysis (e.g. in \cite{Arzoumanian_2018, EPTA_CRS}). The artificially introduced hyperparameter (model index) allows the sampler also to jump between several models and the BF is given by a ratio of the number of samples accumulated in each model.

\subsection{Resampling method}
\label{sec:resampling}

This approach is based on importance sampling and described in detail in \cite{Hourihane:2022ner}. The main idea is to perform sampling like in the ''full'' method neglecting the correlation between pulsars, taking $\Gamma_{ab}=\delta_{ab}$ in \eqref{eq:full_hierar_cov_matrix}. This way the likelihood is factorized into a product of likelihoods for each pulsar \cite{Taylor_2022}. The likelihood computational time is reduced by a factor 6 for six pulsars, but this gain increases as the number of pulsars analyzed increases. Once it is done, we resample (reweigh) the posteriors by recomputing the likelihood for each sample using $\Gamma_{ab}$ defined by the HD curve. As a result, it is possible to parallelize the calculation and therefore gain even more speed. This method works well if the difference in the posteriors obtained with and without HD correlation is not very large (not disjoint distributions).

We can use the same approach to get a cheap estimation of the Bayes factors. To the best of our knowledge, it was not used before, so we provide here some detailed computation.

We refer the reader to section 2 of \cite{Hourihane:2022ner} for details on the resampling of the posterior from an approximate model ``$A$'' to a target posterior ``$T$''. We apply this method to the product space approach for computing the Bayes Factor. We consider two hyper-models $\mathcal{M}$, $\mathcal{M}^\mathrm{HD}$ with the same model parameters $\theta$ and hyperparameter $n$  indexing the models describing the common red noise we want to compare.  
The difference between the two hyper-models is in the likelihood computation:
 we consider uncorrelated common process in the hyper-model $\mathcal{M}$. Computations of Bayes factor with $\mathcal{M}$ are significantly faster and more reliable, we want to evaluate the Bayes factor of models in  $\mathcal{M}^\mathrm{HD}$ by resampling the hyper-chain (product space) $\mathcal{M}$. 
We repeat the steps outlined in section 3 of \cite{Hee_product} but introducing the approximate likelihood in the posterior probability of $n$ for $\mathcal{M}^\mathrm{HD}$
\begin{align}
    P(n | \mathcal{D}, \mathcal{M}^\mathrm{HD}) &= \frac{\pi(n)}{Z_{\mathcal{M}^\mathrm{HD}}} \int \mathcal{L}^{\rm{HD}}(\theta_n)\pi(\theta_n | n) d\theta_n \nonumber \\
    &= \frac{\pi(n)}{Z_{\mathcal{M}^\mathrm{HD}}} \int \frac{\mathcal{L}^{\rm{HD}}(\theta_n)}{\mathcal{L}(\theta_n)} \mathcal{L}(\theta_n) \pi(\theta_n | n) d\theta_n \nonumber \\
    &\approx \frac{\pi(n)}{Z_{\mathcal{M}^\mathrm{HD}}} \frac{1}{N_s^{(n)}}\sum_{i=1}^{N_s^{(n)}}w_i^{(n)} \: Z_n.
\end{align}

Here we introduced $\pi(n)$ the prior on the model index (chosen to be uniform), $\mathcal{L}^{\rm{HD}}$ / $\mathcal{L}$ are likelihood functions with/without HD spatial correlations and $Z_{\mathcal{M}^\mathrm{HD}}$ is the evidence of the hyper-model with HD (in our case just normalization factor which cancels out). In the last line, we have approximated the integral using Monte Carlo approach, with $N_s^{(n)}$ being the number of samples corresponding to the model associated to $n$ in the hyper-model $\mathcal{M}$ posterior. The weights $w_i^{(n)} = \mathcal{L}^{\rm{HD}}(\theta_n^{(i)})/\mathcal{L}(\theta_n^{(i)})$ are computed at each posterior points (indexed by $i$) of $\mathcal{M}$, the $\theta_n$ being the $\theta$ parameters attached to the model $n$.

The Bayes Factor (assuming uniform prior across the models $\pi(n)$) between the models within each hyper-model posterior is given as
\begin{eqnarray}
    B^{ M_a}_{M_b} \equiv B_{a/b} = \frac{P\left(\lfloor n \rfloor = n_a| \mathcal{D}, \mathcal{M}\right)}  {P\left(\lfloor n \rfloor = n_b| \mathcal{D}, \mathcal{M}\right)},
\end{eqnarray}
where $\lfloor . \rfloor$ is the floor function.

Applying this expression to the resampled hyper-model posterior, we obtain
\begin{equation}
    B^{\mathcal{M}^\mathrm{HD}}_{a/b} = \frac{Z_a^{\rm{HD}}}{Z_b^{\rm{HD}}} = \frac{\Bar{w}_{(a)}}{\Bar{w}_{(b)}} \times B^{\mathcal{M}}_{a/b},
\end{equation}
where $\Bar{w}_{(n)} = \frac{1}{N_s^{(n)}}\sum_{i=1}^{N_s^{(n)}}w_i^{(n)}$
is a posterior average weight of the model indexed by $n$. As previously mentioned, the Bayes factor for the hyper-model $\mathcal{M}$ (without HD correlations) is simply given by $B^{\mathcal{M}}_{a/b} \approx N_s^{(n_a)}/N_s^{(n_b)}$. The performance of this method depends on the sampling efficiency, as described in \cite{Hourihane:2022ner}. Note that the novelty of this method is in the efficient evaluation of Bayes factor between \emph{two or more GWB models} via resampling.

\subsection{``Free spectrum'' method}
\label{sec:FSMeth}

As mentioned, this method was suggested in \cite{Moore_2021} and further developed in \cite{Taylor_NFS}. The main idea is to get rid of the high dimensionality of the problem by marginalizing over the single pulsar noise parameters. We want to estimate the PSD assuming the HD correlation at each frequency bin  $f_k = k/T$, where $T$ is the total time span of the array. Using MCMC we get a posterior distribution for the amplitude $\rho_k$  at each frequency  bin, where 
\begin{equation}
    \label{eq: rho expression}
    \rho_k^2 = \frac{S(f_k)}{T},
\end{equation}
where $S(f)$ is the one-sided PSD of the SGWB we want to characterize, see Eq.~\eqref{eq:StoOmegaGW}.   We assume that all bins are independent and impose log-uniform prior for each amplitude, $ \log_{10} \rho_k \in [-12, -4]$, a justification for this range is given in \autoref{sec: model comp}.

We use a kernel density estimator to obtain a smooth probability distribution function $p(\rho_k | \dt)$.  In this study, we use Gaussian kernels, and their bandwidth is selected using Scott's rule of thumb \cite{Scott_book}. We use the estimated PSD of a common process with HD correlation as our new data. We can then build a factorized likelihood that depends only on the parameters of the SGWB $\hyp$,  following \cite{Taylor_NFS}, we get
\begin{align}
    p(\dt | \hyp) &\approx \prod_{k=1}^{N_f} \int \dd{\rho_k} p(\dt | \rho_k) p(\rho_k | \hyp) \notag \\
    &\propto \prod_{k=1}^{N_f} \int \dd{\rho_k} \frac{p(\rho_k | \dt)}{p(\rho_k)} \: p(\rho_k|\hyp).
    \label{eq:FS_Gen_Fact_Lklhd}
\end{align}

We have used the following notation
: $p(\rho_k | \dt)$ are the posterior distributions of the $\rho_k$ describing our observations in Fourier domain, $p(\rho_k)$ is the prior on the $\rho_k$ and $p(\rho_k|\hyp)$ is the probability to have $\rho_k$ given the CRN parameters $\hyp$. For the latter, we consider that the PSD is perfectly deterministic such that 
\begin{equation}
    \label{eq: rho_prob_given_eta}
    p(\rho_k|\hyp) = \delta\left(\rho_k - \rho_k^\mathrm{SGWB}(\hyp)\right),
\end{equation}
where $\rho_k^\mathrm{SGWB}(\hyp)$ is derived from the PSD of a corresponding SGWB model computed at $f_k = k/T$ for a set of background parameters $\hyp$.

Note that the Fourier frequency bins are not entirely independent as the PTA data are not evenly sampled and pulsars do not have the same observational duration. We have computed the associated Pearson correlation coefficient matrix $P_{i,j} = C_{i,j} / \sqrt{C_{i,i} C_{j,j}}$ where $C_{i,j}$ is the covariance matrix of the $\rho$'s. We have found that the frequency bins are approximately independent: the average of the non-diagonal coefficients being $\langle |P_{i,j}| \rangle_{i < j} \approx 0.02$.

Taking all into account, the likelihood of \cref{eq:FS_Gen_Fact_Lklhd} takes the form
\begin{equation}
    \label{eq:FS_part_Fact_Lklhd}
    p(\dt | \hyp) \propto \prod_{k=1}^{N_c} p\left(\rho_k^\mathrm{SGWB}(\hyp) | \dt \right).
\end{equation}

This method is extremely fast and drastically reduces the dimensionality of the problem, making it possible to use the Nested sampling
\cite{nested_sampling_2004, nested_sampling_2006} algorithm to obtain the evidence for each model and posterior distributions of the SGWB parameters $\hyp$.  We used a particular implementation of nested sampling, \texttt{dynesty}, described in \cite{dynesty_code, dynesty_paper}.

\section{\label{sec:results}Results}

Before discussing our results, we should mention that the six pulsars dataset considered here is non-informative about the presence of the HD correlations, as was shown in \cite{EPTA_CRS, Chalumeau_2021}. Consequently, we expect that the results of the analysis will be similar (though not identical) using CURN or GWB assumptions for the CRN. The main purpose of this paper is to introduce the methodology which will be applied to the extended (25-pulsar) EPTA dataset.

We start by considering the cosmic string models with the fixed average number of cosmic string bursts per oscillation to $N_c = 2$ and $N_k = N_{kk} = 0$, leading to $\Gamma = 57$, which is expected by a population of smooth loops \cite{Blanco_Pillado_2015}.
Later we extend our analysis by allowing the number of kinks to vary (and so the value of $\Gamma$).

\subsection{Parameter estimation}

\begin{figure*}
    \includegraphics[width=.75\textwidth]{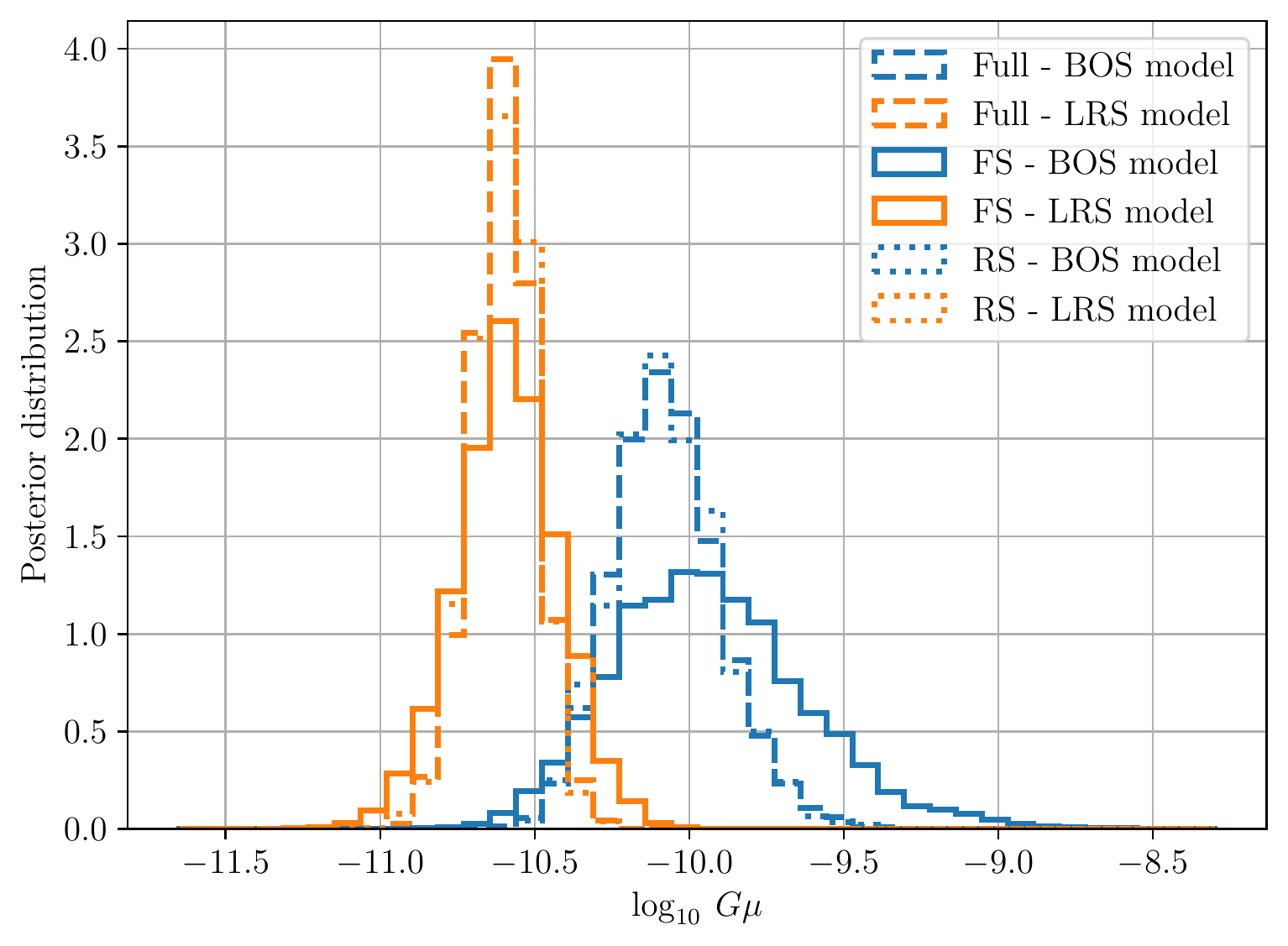}
    \caption{Comparison of the string tension posteriors (for two string models, BOS and LRS) obtained with the Full method (dashed lines), Resampling (RS) method (dotted lines) and with the free spectrum (FS) method (solid lines). We assume here that the loops are populated by two cusps, leading to $\Gamma = 57$.}
    \label{fig: tension_distrib}
\end{figure*}

\subsubsection{Smooth loops}

When we infer a SGWB from a network of smooth loops with $N_k=0$ and $N_c=2$ (alongside the individual pulsar noise models), we find a very constrained distribution for the tension $G\mu$ for both loop density distribution models, see \autoref{fig: tension_distrib}. As one can see  the posterior distributions given by the Full, the RS and the FS methods agree very well, though we find that the FS method gives slightly broader posteriors. The string tension 90\% credible (symmetric) intervals for each Bayesian method can be found in the two first lines of \autoref{tab: post and BFs}. These intervals show nice consistency across all three Bayesian methods: $\logGmu \sim -10.1$ (resp. $-10.6$) for the BOS (resp. LRS) model.

These results can be understood as follows. For the BOS model, the SGWB exhibits a peak around $f \sim F_0 \equiv 3 H_0 (\Gamma G \mu)^{-1}$ before decreasing to reach a plateau at higher frequencies (see \autoref{fig:SGWB_wide}) \cite{Auclair:2020oww} . 
According to the posterior for $G\mu$, the peak is in the middle of the PTA band, thus the spectral index of $\Omega_{\rm{SGWB}}$ transits from 1 to 0 (flat spectrum) throughout the PTA frequency range. This is comparable to the value of 2/3 (corresponding to $\gamma = 13/3$) expected for a SMBHB background. As can be seen in \autoref{fig: FS_plot}, the BOS model slope is very compatible with the spectral estimated from the data.

Regarding the LRS model, the extra population of small loops (see \cite{Auclair:2020oww}) is responsible for the fact that instead of decreasing to a plateau for $f > F_0$, the characteristic energy density rather increases again up to a second peak at very high frequencies compared to the PTA band, see \autoref{fig:SGWB_wide}. The fractional energy density follows there a power-law of spectral index $\approx 0.4$, which also fits well the spectral shape of the data (see \autoref{fig: FS_plot}).

As an alternative, we also consider a two-component SGWB model: a SGWB generated by a population of circular binary black holes with $\gamma = 13/3$ plus a stochastic signal from the cosmic strings.

\begin{figure*}
    \includegraphics[width=.49\textwidth]{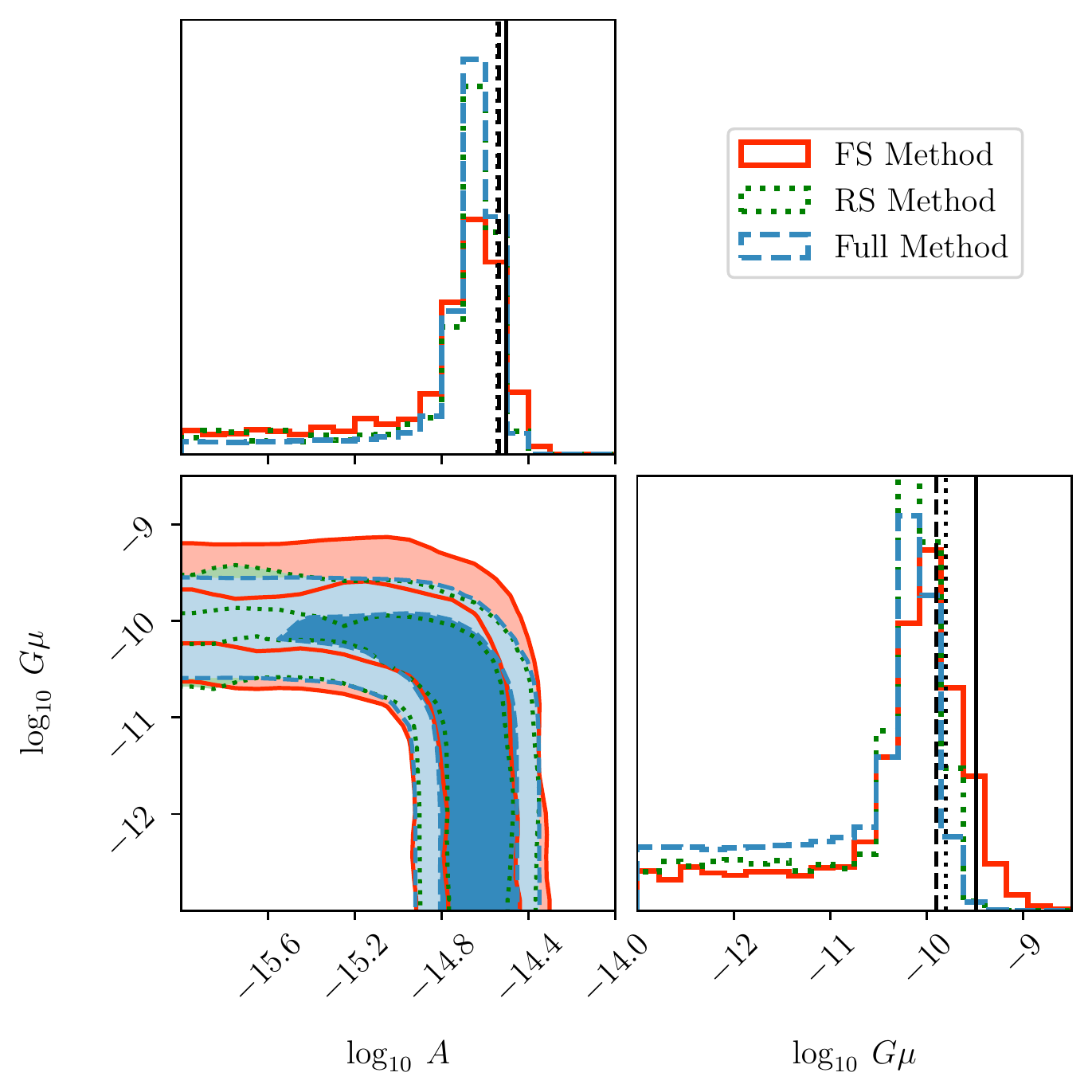}
    \includegraphics[width=.49\textwidth]{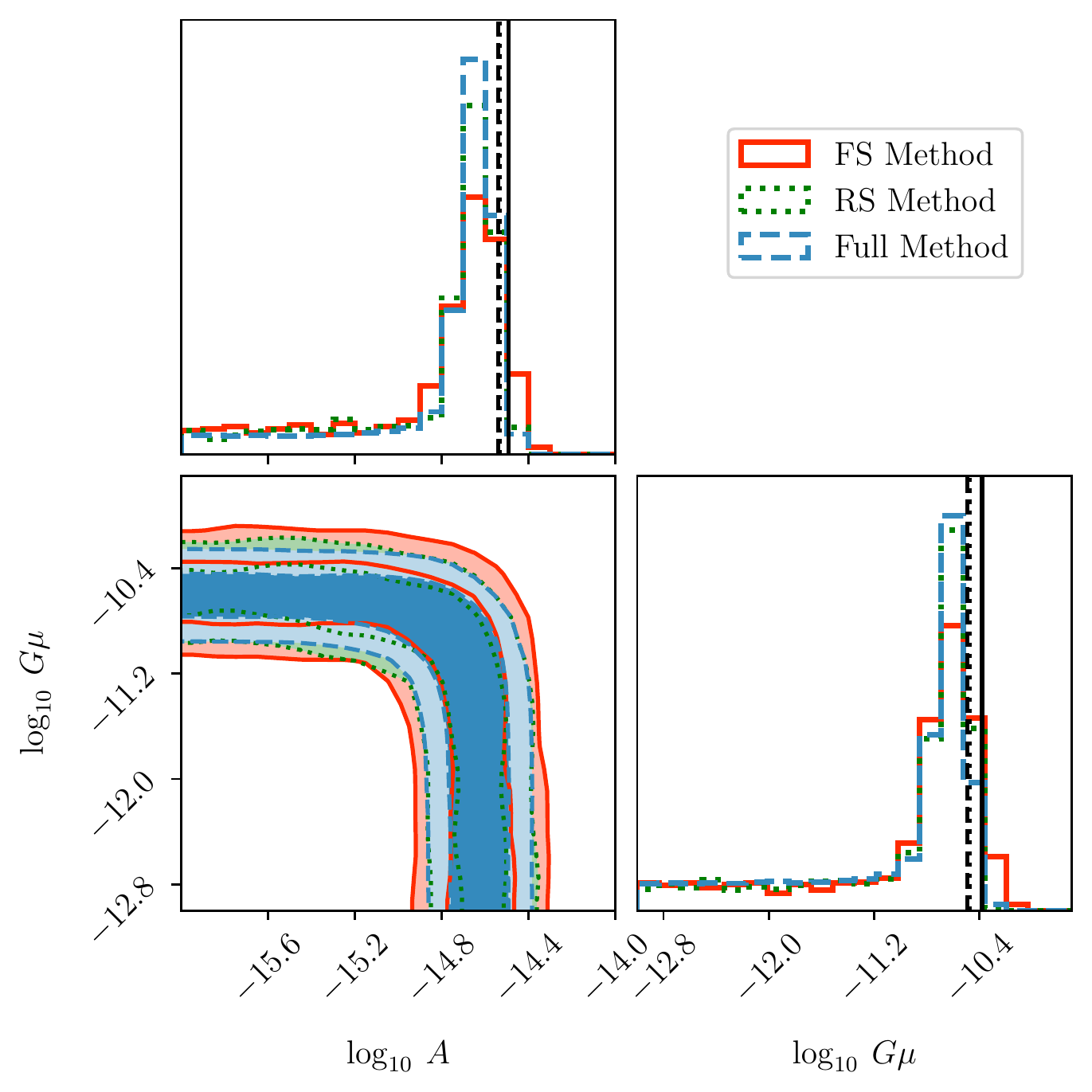}
    \caption{
        Left panel: posterior for the two-component SGWB model composed of (i) a signal originating from a population of circular GW-driven SMBHB, parameterised by its PSD amplitude $\log_{10} A$ at $f=1/$year, and (ii) a SGWB from smooth CS loops background using the BOS loop number density model; different lines styles (dashed, dotted, solid) corresponding to three methods (Full, RS, FS)  show good consistency. The $95$-quantile for each SGWB parameter posterior is plotted in black, using the line style associated with its respective method. Right panel: the same for the LRS loop number density model.}
    \label{fig:2d-CS-BBH-post}
\end{figure*}

We find a very strong correlation between those two components as one can see in \autoref{fig:2d-CS-BBH-post}. The 2D-posteriors exhibit a pronounced L-shape, meaning that both model can explain the data. Based on the obtained posterior distribution, we can extract an upper limit which we evaluate to be $\logGmu < -9.9$ (resp.~$-10.5$) for the BOS (resp.~LRS) model using the Full method, and it  corresponds to 95\% confidence. In addition, the evaluation obtained with the two fast methods (RS, [FS])  give very consistent upper bounds: $\logGmu < -9.8 [-9.5]$ (resp.~$-10.5 [-10.4]$) for the BOS (resp. LRS model).
The slightly broader posterior obtained with FS earlier leads to a less stringent upper bound for the BOS model.
One can notice that we constrain better the tension, $G\mu$, for the LRS model, this is due to the excess of power that the small loop population introduces.

\subsubsection{Kinky loops}

Next we also vary the number of kinks (and thus kink-kink collisions) in addition to the string tension, considering two-dimensional models  $(\logGmu, N_k)$.   The PSD  of the SGWB created by such a model is calculated from \cref{eq: full SGWB energy density} by setting $N_c = 1$ (in order to still have some GW power at the low number of kinks) and $N_{kk} = N_k^2/4$. 

\begin{figure*}
    \includegraphics[width=.49\textwidth]{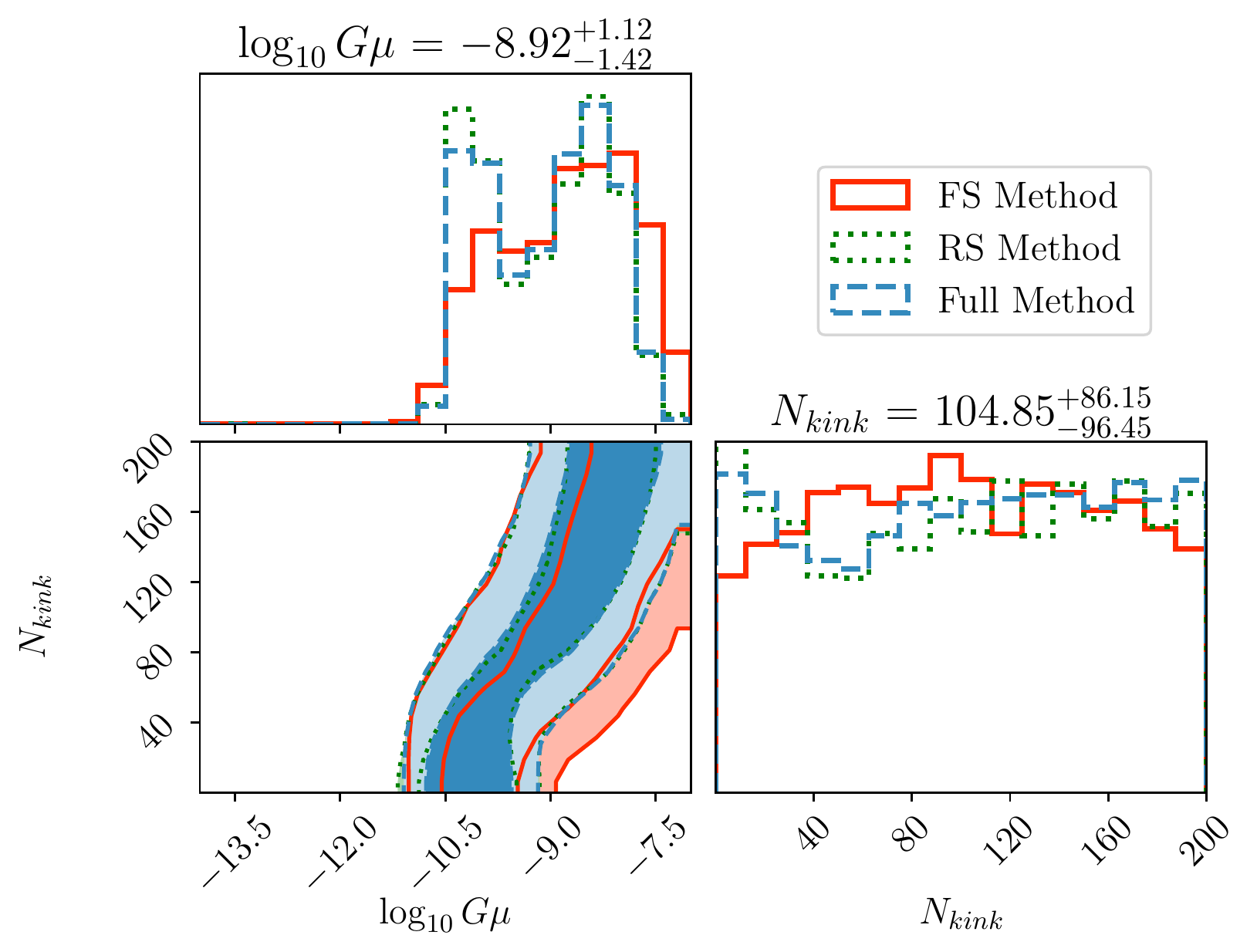}
    \includegraphics[width=.49\textwidth]{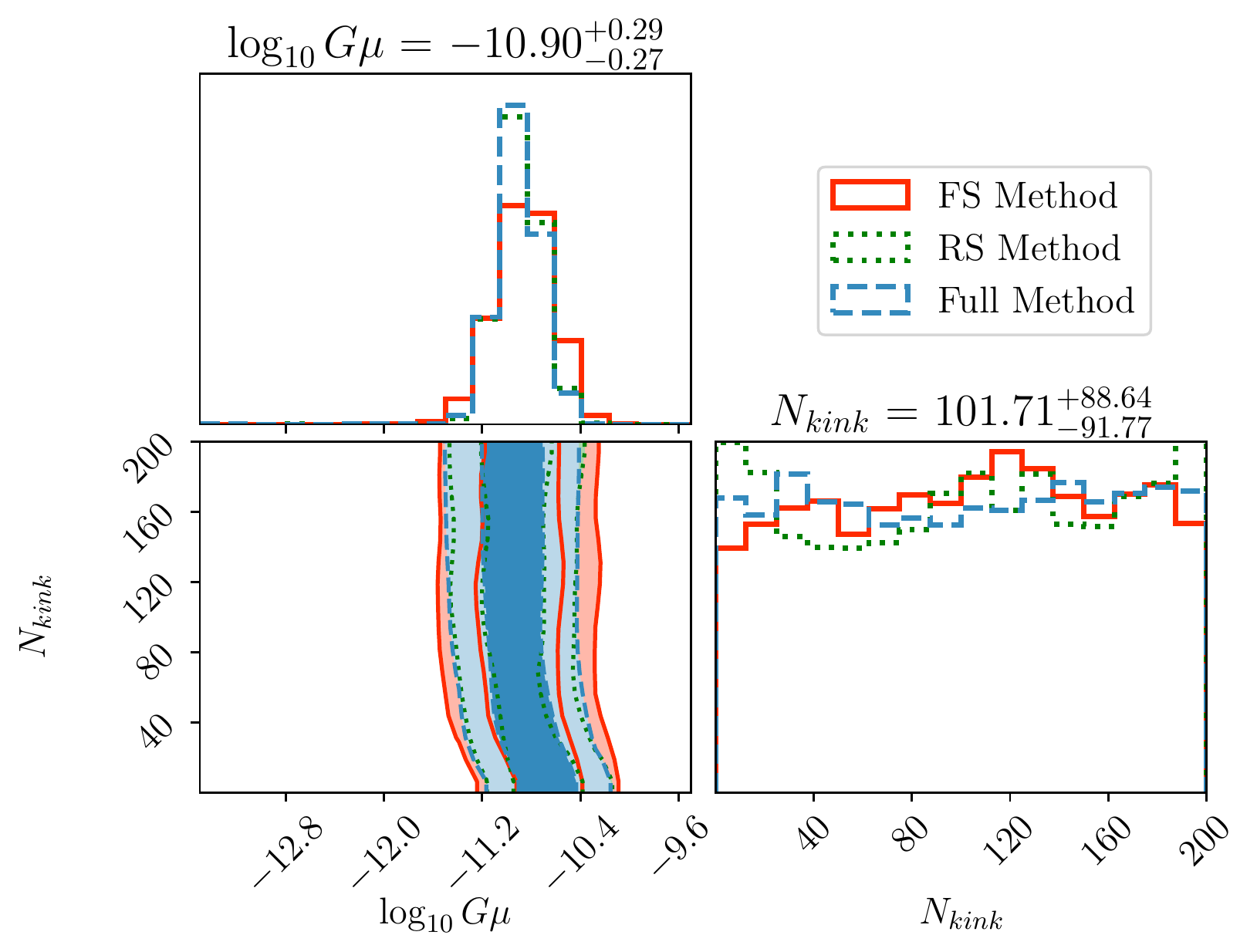}
    \caption{
        Left panel: posterior for two-dimensional (string tension and the average number of kinks on loops of the network)  BOS model; different lines styles corresponding to three methods (Full, RS, FS)  show full consistency.  Right panel: the same for the LRS loop number density model.
    }
    \label{fig:2d-CS-post}
\end{figure*}

We took a uniform prior for $N_k$ letting it vary between $0$ and $200$,  the same prior was used in the LVK analysis \cite{LIGOScientific:2021nrg}. This large prior on $N_k$ accounts for  theoretical uncertainties on the initial number of kinks at loop creation and on the efficiency of the gravitational backreaction that is expected to smooth out the loop.  

In the case of the BOS model, if we increase the number of kinks (and so $\Gamma$), the SGWB in the PTA band now corresponds to the transition between the peak of the spectrum and the high frequency plateau and thus increases the characteristic slope of its PSD while decreasing the amplitude. This increase in spectral index will be disfavored as it will be too steep to fit the data properly. However, to compensate the decrease in amplitude, an increase of the string tension $G\mu$ can both correct the amplitude and the spectral index by placing the high frequency plateau in the PTA frequency band.
 One can see such interplay between $G\mu$ and $N_k$ in the left panel of \autoref{fig:2d-CS-post}. The data equally allows two joined solutions: 
 low tension with low number of kinks and very kinky loops ($N_k \gtrsim 120$) with high tension, $\logGmu \sim -8.3$.

 Contrary to the BOS case, the PSD slope of the LRS increases only slightly at low frequency with increasing $N_k$. Once $F_0 \lesssim 1/T \sim 10^{-9}$, the spectrum is dominated by the extra population of small loops that the LRS model introduces and the associated power is now independent of $\Gamma$ and proportional to $G\mu$ \cite{Auclair_2020}. Therefore, the only shift in amplitude is caused by the domination of kinks (instead of cusp) in the GW power emission leading to an increase by a factor $\sim 1.6$ \cite{Auclair_2020}, that can be simply compensated by a slight decrease in the tension leading to $\logGmu \sim 10.9$. As a result, adding kinks has practically no effect in the LRS model, and we recover the prior for $N_k$, as seen  the right panel of \autoref{fig:2d-CS-post}.

According to the posterior, the model with a high number of kinks is supported by the observed data, especially for BOS where it suggests a possible 
 very high number of kinks. For example, $N_k = 120$ corresponds to $\Gamma \approx 960$,  which is well above the value of 50 expected in the latest simulations of cosmic string loops. Finally, all three methods (depicted with different line styles in \autoref{fig:2d-CS-post}) show very consistent results, even if the FS method struggles more to recover the double peak solution in the BOS model case.

\subsection{\label{sec: model comp}Model comparison}

\begin{figure*}
	\includegraphics[width=.7\textwidth]{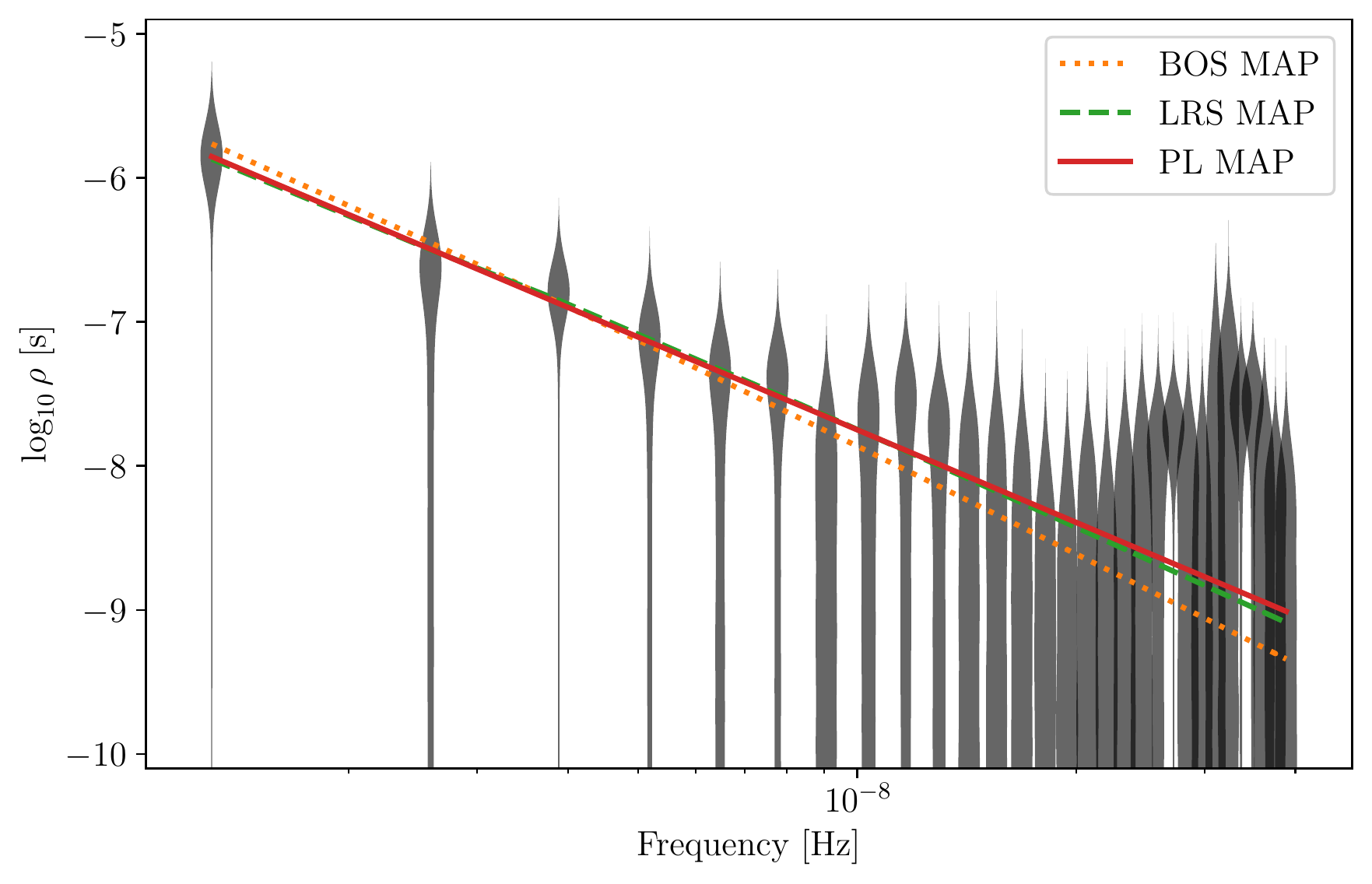}
	\caption{Posterior distributions of the 30 $\rho_k$ coefficients (in $\log_{10}$-scale). We over-plotted the best fit (using the FS likelihood of \cref{eq:FS_part_Fact_Lklhd}) for three different PSDs: powerlaw, SGWB using BOS and LRS models (in the case of smoothed loops). We see that all three spectra behave in a similar way at low frequency bins.}
    \label{fig: FS_plot}
\end{figure*}

The fact that the cosmic string parameters are well constrained  does not tell us how well it explains the observed data.  The Bayes factor, defined as the ratio of evidence between two models,  is often used to quantify  the ability of two different models to fit a set of data. Note that the Bayes factor can tell us which model is preferred based on the observations, but still does not tell us how well it describes the data. In what follows, we want to compute the Bayes factors of the considered cosmic string models against the power-law (PL)  (SMBHB-inspired) model.  Results are reported in \autoref{tab: post and BFs}. We find that the cosmic string models are slightly preferred over PL model thanks to its simplicity: only one parameter. However, \autoref{fig: FS_plot} shows that all models fit the spectral shape similarly well.

Considering the two-component model of SGWB (SMBHB+CS, see previous subsection), we have found no statistical support for this complex model as compared to a simple SMBHB, $\mathcal{B}^\mathrm{SMBHB+CS}_\mathrm{SMBHB} \approx 1$ for both BOS and LRS cosmic string networks (using Full, RS and FS methods).

We have demonstrated earlier that the kinks had no strong impact and were not constrained by the data in the models with the variable number of kinks. This is also reflected in the evaluated Bayes Factors.

\begin{figure}
	\includegraphics[width=.4\textwidth]{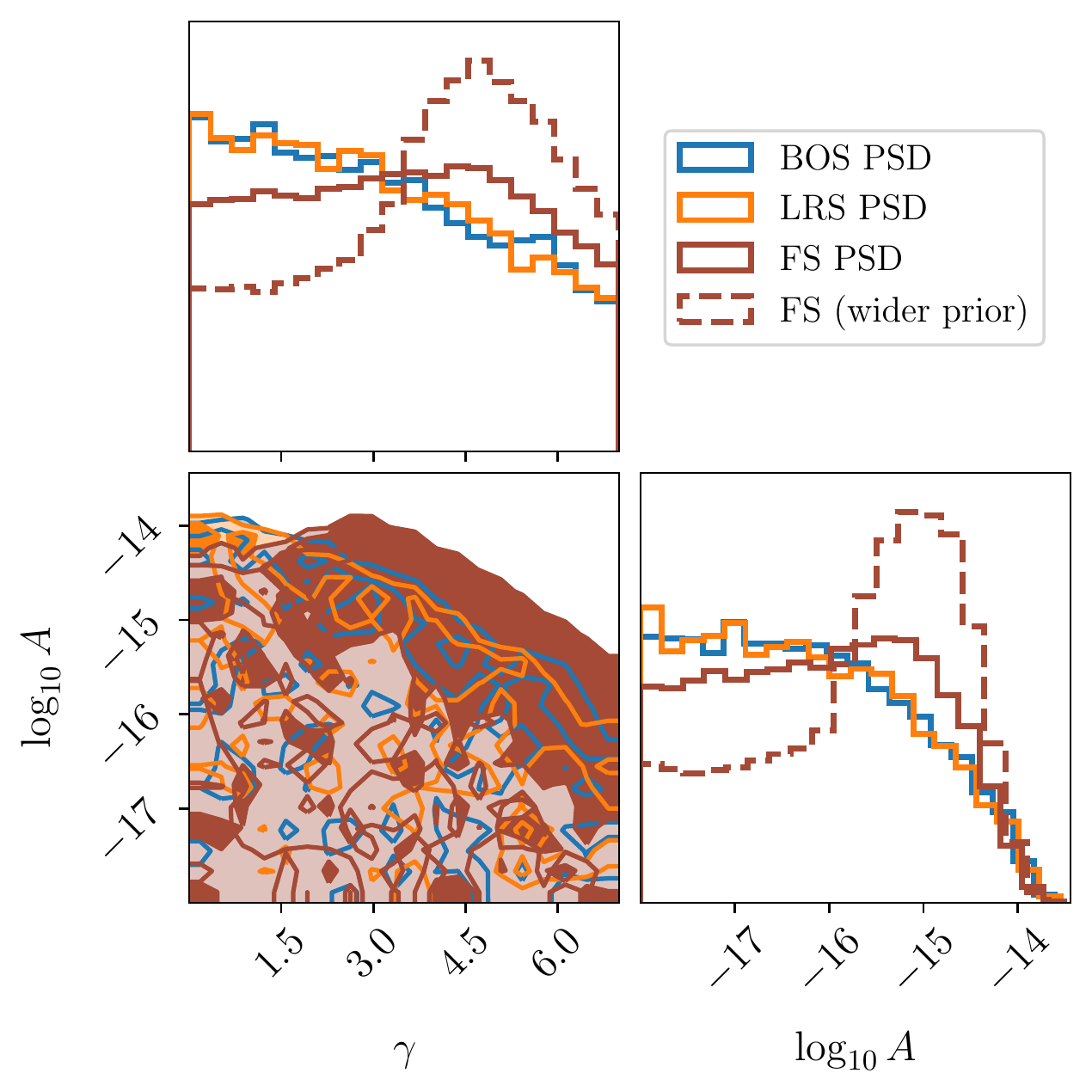}
	\caption{Marginalized red noise parameters posteriors for the pulsar J1909-3744 obtained with the Full method including HD correlation for the common red noise process.  Brown dashed and solid lines were obtained using different prior on the power in the Fourier bins (passing from $-10$ to $-15$ for the $\log_{10}\rho_k$ lower bound). The solid line (restricted prior) suggests a truncated correlation of the common red noise with the spin red noise of J1909-3744.
        We also plotted for comparison those posteriors when using as common red noise, a SGWB from cosmic string following the BOS (blue line)/LRS (orange line) models.
		}
	\label{fig: 1909_post}
\end{figure}

Both Full and the RS methods give very consistent Bayes factors for all models, as shown in \autoref{tab: post and BFs}.

The Bayes factors computed with FS method depend on the prior chosen for evaluating the FS amplitude components $\rho_k$. The main reason is that power at all low frequency bins is not fully constrained. The posteriors have a long tail towards the low values of power (see \autoref{fig: FS_plot}). Choosing a wider prior makes those tails longer, spreading the probability across a long range of $\rho_k$ (draining it from the "bumps" located at high power). Despite the fact that the tails are thin, they go all the way to minus infinity in log-scale (zero in power) and the probability distribution is then strongly affected (through normalization). So we are facing the question of what we should use as a prior range for the $\log_{10} {\rho_k}$?

The choice of the prior range has a direct impact on the evidence calculation: some values of the SGWB model parameters have no support (zero contribution to the evidence integral) because of the truncation of the $\rho_k$ imposed by the prior (which is not the case in the Full or RS approach). Even more important is the indirect impact caused by the correlation of pulsar spin noise (red noise intrinsic to each pulsar) and common red noise at the lowest frequencies. In \autoref{fig: 1909_post}, we show the inferred distribution for the parameters of the spin red noise in one of the best pulsars, PSR J1909-3744. Extending the prior range for $\rho_k$ leads to steeper and stronger spin red noise in that pulsar (dashed brown line  compared to the solid brown line). This implies that the (artificially) over-constrained  prior on $\rho_k$ truncates the noise estimation in the  pulsar J1909-3744. 
We empirically determined a fiducial value of power in terms of $\log_{10} \rho_{th}$ (that can be roughly regarded as common to all frequency bins) which represents the threshold below which our data becomes insensitive. We selected a symmetric range centered around this specific value, which we set at $\log_{10} \rho_{th} = -8$. For the upper limit, we opted for $-4$, leading to a lower bound of $-12$ for the $\log_{10} \rho$ prior. This prior answers direct (albeit partially) and indirect (by preserving correlations) impact on the Bayes factor while avoiding an overblow of the prior range.

In addition, it is worth noting \footnote{We thank the referee for pointing us in this direction.} that the posteriors given by the FS method typically have a broader spread compared to other methods. Moreover, the FS method encounters difficulties in accurately capturing the double peak feature observed in the $G\mu$ posterior for the BOS model in \autoref{fig:2d-CS-post}. This issue is mainly due to the width of the prior for the $\rho_k$ parameters. Indeed, if a $\rho_k$ posterior is not well-constrained, which is the case for several frequency bins as evident in \autoref{fig: FS_plot}, taking a wider prior results in a longer tail and less constrained posterior at its upper bound (due to normalization).
 This gives a likelihood function  (\autoref{eq:FS_part_Fact_Lklhd}) with less pronounced peaks, leading to the spread in the posterior distributions.
By running the FS method using a narrower prior range of $[-10, -4]$ for the $\rho_k$, we recover the posterior distribution which is much closer to the one obtained with the Full method.

In general, we have found the choice for the prior for the free spectrum evaluation (or rather a lack of rigorously defined cut) is a weak point of the FS approach in computing the evidence, despite that it perfectly falls into Bayesian philosophy. 

Accumulating high-quality data and using more pulsars in the array, will hopefully result in the well-constrained power at low frequencies. Consequently, this will make the choice of the prior for the $\rho_k$ parameters much less relevant for evaluation of the Bayes factors using the FS method.

\begin{table*}
  \centering
  \renewcommand{\arraystretch}{1.2}
  \begin{ruledtabular}
      \begin{tabular}{p{2.5cm}cccccc}
        \multirow{2}{2.5cm}{} & \multicolumn{2}{c}{\textbf{Full Method}} & \multicolumn{2}{c}{\textbf{RS Method}} & \multicolumn{2}{c}{\textbf{FS Method}}\\

        \cline{2-7} & \textbf{CS Posterior} & \textbf{BF (PL/CS)} & \textbf{CS Posterior} & \textbf{BF (PL/CS)} & \textbf{CS Posterior} & \textbf{BF (PL/CS)}\\[4pt]

        \textbf{BOS} & $\log_{10} G\mu = -10.08^{+0.32}_{-0.26}$ & 0.3 & $-10.08^{+0.30}_{-0.27}$ & 0.2 & $-9.95^{+0.56}_{-0.45}$ & 0.1 \\[7pt] 

        \textbf{LRS} & $\log_{10} G\mu = -10.60^{+0.17}_{-0.17}$ & 0.2 & $-10.60^{+0.16}_{-0.18}$ & 0.3 & $-10.59^{+0.27}_{-0.28}$ & 0.2 \\[7pt] 

        \textbf{BOS\_kk} & \begin{tabular}{@{}c@{}}$\log_{10} G\mu = -8.92^{+1.12}_{-1.42}$ \\ $N_k = 105^{+86}_{-96}$\end{tabular} & 0.2 & \begin{tabular}{@{}c@{}}$-8.98^{+1.17}_{-1.38}$ \\ $101^{+90}_{-96}$\end{tabular} & 0.3  & \begin{tabular}{@{}c@{}}$-8.66^{+1.24}_{-1.65}$ \\ $100^{+88}_{-87}$\end{tabular} & 0.1 \\[10pt] 

        \textbf{LRS\_kk} & \begin{tabular}{@{}c@{}}$\log_{10} G\mu = -10.90^{+0.29}_{-0.27}$ \\ $N_k = 102^{+89}_{-92}$\end{tabular} & 0.2 & \begin{tabular}{@{}c@{}}$-10.89^{+0.29}_{-0.27}$ \\ $103^{+89}_{-95}$\end{tabular} & 0.3   & \begin{tabular}{@{}c@{}}$-10.86^{+0.35}_{-0.40}$ \\ $104^{+85}_{-92}$ \end{tabular} & 0.2 \\ 
      \end{tabular}
  \end{ruledtabular}
  \caption{For each Bayesian analysis method (columns), we write the cosmic string parameters posteriors for each loop distribution (BOS/LRS) with a varying number of kinks (\_kk) or without. The 5\% and 95\% quantiles are used to set up the credible interval. The second column quotes the linear Bayes Factors comparing each of the cosmic string (CS) models against a power-law (PL) PSD for the SGWB.}
  \label{tab: post and BFs}
\end{table*}

\section{Conclusion and discussion}
\label{sec:concl}

In this work, we demonstrated that the Free Spectrum and Resampling methods are very powerful tools to determine rapidly the  posterior distributions for the parameters describing the SGWB. Moreover, those methods (taking into account caveats for FS method discussed above) could be used to evaluate the Bayes factor between various SGWB models. We have demonstrated these methods using the 6-pulsar early DR2 EPTA dataset and comparing several models of SGWB produced by a network of cosmic strings. 

For this 6-pulsars only dataset, we could perform rigorous inference (Full method) of  SGWB parameters (in addition to pulsar noise parameters) and Bayes factors evaluation \autoref{tab: post and BFs}. Note that the Full method took several days to complete, whereas the execution time for RS/FS methods is reduced to hours/minutes. 
Extending the data by including more pulsars, more back-ends and new observations will hugely explode the dimensionality of the problem and make the Full method computationally not tractable. The computational power scaling quadratically with the number of pulsars and the number of observations.  We hope that this work convinces the reader of the validity of the fast and approximate methods which have to be used in the future PTA data analysis. 
The attractiveness of FS method is that it is not very sensitive to the number of pulsars in the array. Of course the evaluation of the free spectrum does heavily depend on the volume of the data and dimensionality of  parameters space (pulsars noise models), but it can then be cheaply used to infer parameters for multiple SGWB models. The Resampling method is also relatively cheap, the sampling step scales linearly with number of pulsars in the array and can be efficiently parallelized (if needed).
Even though we have demonstrated the fast methods using simplistic power-law model for SMBHB and two cosmic string SGWB models, the methods are generic and can be used to infer parameters of \emph{any} SGWB. 

The second result of this paper are the constraints on the cosmic string tension $G\mu$ for the BOS and LRS models. Assuming that the observed red noise process is GWs, we have obtained the preferred value for the string tension $G\mu \approx 10^{-10.1}$  (resp.~$10^{-10.6}$) for BOS (resp.~LRS). Moreover, we find that the power-law model with HD correlations is slightly disfavored compared to the simpler one-parameter cosmic strings model. The posterior is not informative about the number of kinks when we consider the two-parameter $(G\mu,N_k)$ cosmic string models.  However, it also implies that the possibility of having a large number of kinks is not ruled out, though it might be disfavored on the theoretical grounds. Considering a two-component SGWB model (SMBHB+CS) shows a strong correlation between two components, and  we set the upper bounds   $\logGmu \lesssim -9.9$ (BOS model) and  $\logGmu \lesssim -10.5$ at 95\% (LRS model) confidence.

Compared to previous EPTA constraints \cite{Lentati_2015, Sanidas_2012} on $\logGmu$ which were obtained using the EPTA sensitivity curve, here we use more up to date models and consider directly the PSD of the cosmic string SGWB through a wide range of frequencies to obtain our constraints. Our upper bounds on the string tension  are well under the one given by CMB experiment $\logGmu \lesssim -7$ \cite{Charnock_2016} and the LVK collaboration \cite{LIGOScientific:2021nrg} for the BOS model ($\logGmu \lesssim -8$). However, for the LRS model, due to the population of small loops emitting at higher frequencies, the LVK constraint is more stringent,  $\logGmu \lesssim -14$.

This paper lays the path for the search and interpretation of a SGWB signal of any origin. The fast methods suggested here will be applied to the extended EPTA dataset containing 25 pulsars \cite{EPTA:2023fyk}. 
In addition to EPTA data, the new dataset which combines the most sensitive observations from all PTAs, International PTA data combination, is being produced. We expect the future data  to be more sensitive to GWs and give better constraints on the string tension in BOS and LRS models.

\begin{acknowledgments}
H.Q-L, P.A. and D.S. would like to thank the University of Geneva for hospitality while this work was in progress. We are grateful to Chiara Caprini for very useful discussions.
H.Q-L thanks Institut Polytechnique de Paris for funding his PhD.
The work of P.A. is supported by the Wallonia-Brussels Federation Grant ARC \textnumero~19/24-103. S.B. and H.Q-L acknowledge support from ANR-21-CE31-0026, project MBH\_waves.
D.S. is grateful to CERN for hospitality.

\end{acknowledgments}

\clearpage


\bibliography{biblio}
\bibliographystyle{apsrev4-1} 

\end{document}